\documentclass[useAMS,usenatbib]{mn2e}
\usepackage{graphicx}
\usepackage{float}
\usepackage{color}
\usepackage{xspace}

\usepackage{color,multirow}
\usepackage{lscape}
\usepackage{soul}

\usepackage[backref,breaklinks,colorlinks,citecolor=blue]{hyperref}
\usepackage[all]{hypcap}
\usepackage{amsmath}	% Advanced maths commands
\usepackage{amssymb}	% Extra maths symbols
\usepackage{xspace}
 
\usepackage[normalem]{ulem}
\usepackage{soul}

\DeclareRobustCommand{\ion}[2]{%
\relax\ifmmode
\ifx\testbx\f@series
{\mathbf{#1\,\mathsc{#2}}}\else
{\mathrm{#1\,\mathsc{#2}}}\fi
\else\textup{#1\,{\mdseries\textsc{#2}}}%
\fi}

\def\arcsec{\hbox{$^{\prime\prime}$}\xspace}
\def\arcmin{\hbox{$^{\prime}$}\xspace}

\newcommand{\Msunyr}{\hbox{$M_\odot \,\hbox{yr}^{-1}$}\xspace}

\newcommand{\rs}{$R_{\odot}$\xspace}
\newcommand{\ls}{$L_{\odot}$\xspace}
\newcommand{\kms}{km\,s$^{-1}$\xspace}

\newcommand{\ec}{$\eta$\,Car\xspace}

\newcommand{\ha}{H$\alpha$\xspace}
\newcommand{\hd}{H$\delta$\xspace}

\title[coronagraph]{Spectroscopic Signatures of the Vanishing Natural Coronagraph of eta Carinae\\}

\author[Damineli et al.]
  {A.~Damineli$^{1}$\thanks{E-mail: augusto.damineli@iag.usp.br},
     F.~Navarete$^{1}$,
    D.~J.~Hillier$^{2}$,
      A.~F.~J.~Moffat$^{5}$,
    M.~F.~Corcoran$^{9,10}$,
     \newauthor 
     T.~R.~Gull$^{3}$,
      N.~D.~Richardson$^{4}$,
    G.~Weigelt$^{6}$, 
    P.~W.~Morris$^{7}$,
    I.~Stevens$^{8}$,\\
  $^1$ Instituto de Astronomia, Geof\'isica e Ci\^encias Atmosf\'ericas da USP, Rua do Mat\~ao 1226, Cidade Universit\'aria, S\~ao Paulo, Brasil \\
  $^2$ Department of Physics and Astronomy \& Pittsburgh Particle Physics, Astrophysics, and Cosmology Center (PITT PACC), \\ \hspace{1cm}  University of Pittsburgh, 3941 O'Hara Street,  Pittsburgh, PA 15260, USA\\
$^3$ Laboratory for Extraterrestrial Planets and Stellar Astrophysics, Code 667, NASA Goddard Space Flight Center, Greenbelt, MD 20771, USA \\
$^4$ Department of Physics and Astronomy, Embry-Riddle Aeronautical University, 3700 Willow Creek Road, Prescott, AZ 86301, USA \\
$^5$ D\'epartement de Physique and Centre de Recherche en Astrophysique du Qu\'ebec (CRAQ)\\ Universit\'e de Montr\'eal, C.P. 6128, Succ. Centre-Ville, Montr\'eal, Qu\'ebec, H3C 3J7, Canada \\
$^6$ Max Planck Institute for Radio Astronomy, Auf dem H\"{u}gel 69, D-53121 Bonn, Germany \\
$^7$ California Institute of Technology, IPAC M/C 100-22, Pasadena, CA 91125, USA\\
$^8$  School of Physics \& Astronomy, Univ. of Birmingham, Edgbaston, Birmingham, B15 2TT, UK\\
 $^{9}$ CRESST II \& X-ray Astrophysics Laboratory, Code 662, NASA Goddard Space Flight Center, Greenbelt, MD 20771, USA\\
  $^{10}$ Institute for Astrophysics and Computational Sciences, Department of Physics \\
  The Catholic University of America, Washington, DC 20064, USA} 

\date{May 10 2021}

\begin{document}
\label{firstpage}

\maketitle
\begin{abstract} 
Eta Carinae is a massive interacting binary system shrouded in a complex circumstellar environment whose evolution is the source of the long-term brightening observed during the last 80 years. An occulter, acting as a natural coronagraph, impacts observations from our perspective, but not from most other directions. Other sight-lines are visible to us through studies of the Homunculus reflection nebula. The coronagraph appears to be vanishing, decreasing the extinction towards the central star, and causing the star's secular brightening. In contrast, the Homunculus remains at an almost constant brightness. The coronagraph primarily suppresses the stellar continuum, to a lesser extent the wind lines, and not the circumstellar emission lines. This explains why the absolute values of equivalent widths (EWs) of the emission lines in our direct view are larger than those seen in reflected by the Homunculus, why the direct view absolute EWs are decreasing with time,  and why lower-excitation spectral wind lines formed at larger radii (e.g \ion{Fe}{ii} 4585{\AA}) decrease in intensity at a faster pace than higher excitation lines that form closer to the star (e.g. \hd). Our main result is that the star, despite its 10-fold brightening over two decades, is relatively stable. A vanishing coronagraph that can explain both the large flux evolution and the much weaker spectral evolution. This is contrary to suggestions that the long-term variability is intrinsic to the primary star that is still recovering from the Great Eruption with a decreasing mass-loss rate and a polar wind that is evolving at a slower pace than at the equator.
\end{abstract}

\begin{keywords}
ISM: dust, extinction ISM: molecules -- stars: binaries -- stars: individual: eta Carinae -- stars: winds
\end{keywords}

\newpage

\section{Introduction}
\label{sectionintroduction}

The evolution and fate of massive stars are often determined by binary interaction \citep[e.g.][]{sana12}. Their scarcity in the solar neighbourhood makes progress in the field difficult. \ec is relevant because of its proximity (2.3\,kpc), its long observational history, and because it is one of the most massive stars within our Galaxy. It owes its fame to the great eruption (GE) that occurred in the 1840s, when it became the second brightest star in the night sky and ejected the elegant bipolar Homunculus nebula \citep{humphreys97}. Hidden behind circumstellar dust \citep{andriesse1978, viotti1989}, the central object is a highly-eccentric massive binary \citep{damineli96, damineli97,damineli00} with  periastron passages occurring every 5.538\,yr and that are monitored throughout the electromagnetic spectrum  \citep{pittard02,Corcoran17,Duncan2003,Lajus2010,Mehner2014,damineli08,teodoro16}. 

The primary star (\ec\,A) dominates the optical light of the system. Its extinction-corrected apparent magnitude would be $V$\,=\,0.94\,mag, according to modelling of the optical/UV spectrum \citep{hillier06}, plus a foreground extinction of up to $A_{\rm V}$\,=\,2.5\,mag \citep{teodoro20}. The apparent magnitude of the central stellar core was $V$\,=\,7.73 on 1999.140 \citep{martin04}, indicating additional local extinction obscuring our view. If the dust was uniformly distributed around the stellar core, the Homunculus reflection nebula would be proportionately fainter. The simple fact that the Homunculus is so prominent, compared to the central star,  indicates extra extinction in our line of sight (LOS) to the central star. Moreover, the relative high brightness of the three speckle objects BCD \citep{weigelt86} could be explained if they are illuminated by the central star and seen in an unimpeded view away from our LOS, while the central star, located a fraction of an arcsec away, is seen under heavy extinction caused, for example, by an obscuring dusty disk \citep{weigelt95} or some other optically thick, localized structure.

From the 1940s through the 1990s, the \ec\ system (star + Homunculus) had continuously brightened at a rate of $\delta V$\,$\sim$\,$-$\,0.02\,mag\,yr$^{-1}$ in the $V$-band. However, existing imagery from those epochs cannot separate the stellar contribution from the nebula. Over the past two decades, the brightening increased to a rate of $\delta V$\,$\sim$\,$-$0.05\,mag\,year$^{-1}$ \citep{martin06}, and improved photometric imagery allowed the separation of the stellar core from the flux of the Homunculus nebula: the stellar core  brightened at a rate $\delta V$\,$\sim$\,$-$0.11\,mag\,yr$^{-1}$ while the Homunculus nebula, has remained fairly constant at $V$\,=\,5.48$\,\pm$\,0.05\,mag \citep{damineli19}. The light emitted by the central star and reflected in the NW extremity of the Homunculus reaches the Earth with a delay of  $\sim$\,6\,months.  If the star was intrinsically brightening, the Homunculus light-curve should show the same effect delayed by a few months. 
In 2010 the brightness of the central star surpassed that of the Homunculus in the $V$-band, for the first time in 160 years. Since the Homunculus is a reflection nebula and is almost constant in brightness, the extinction along our LOS to the central star must be changing \citep{damineli19}. The brightening cannot be explained by the central star becoming intrinsically more luminous since the stellar spectrum has not evolved significantly even though the object brightened by approximately a factor of 10 in the last 70 years. No such long-term brightening was observed in the mid-infrared, as reported by \citet{mehner19}. The sparse observations in the mid-infrared since 1968 are compatible with lower amplitude variations (factor $\sim$\,2) on a timescale of years, as is the case also for the {\it orbital modulation} in the optical reported by \citet{damineli19}. The mid-infrared flux is produced by the absorption and re-emission of light from the central object by dust spread over the whole nebula. In contrast to the optical and ultraviolet emission, it is extended, and not affected by any localised structure.

{\ec\ is a strong  X-ray source, in which thermal X-rays are produced in the hot shocked gas where the wind from the primary star collides with the wind of the companion.  The luminosity and temperature of the X-ray emission depend on the wind properties (i.e. mass-loss rates and pre-shock wind velocities: see for example \citet{stevens92}. The wind parameters in turn depend on the underlying stellar parameters.  Thus, variations in the intrinsic stellar properties would be reflected in variations in the observed X-ray spectrum.  The X-ray temperature and luminosity near apastron (when Coriolis distortions are minimal) show little variations from orbit-to-orbit over the last four orbits for which detailed monitoring data exist - see \citet{Corcoran17} and \citet{espinoza21} - indicating that neither the wind nor stellar parameters have varied significantly during the prior two decades. This conclusion  is supported  by the lack of variations in the terminal velocities derived from the P-Cygni absorption profiles of the  upper members of the Balmer series (like \hd ) \citep{teodoro12}.}

{The initial suggestion that we see the central star obscured by dust condensation and that the Homunculus scatters radiation that has suffered relatively little attenuation was made by \citet{hillier92}. This idea arose because the EWs of the lines seen in direct light from the central star were larger than the corresponding EWs measured in reflection from the Homunculus. This was true for some wind lines (\ion{Fe}{ii}, \ion{H}{i}), and particularly true for the narrow nebular lines. Indeed, the reflected spectrum was much more similar to a classic P\,Cygni spectrum than the spectrum of \ec\ observed in direct light. \citet{hillier06} discussed additional evidence of the presence of a kind of natural coronagraph in \ec.}

{Based on spatially-resolved HST/STIS spectroscopy of the star, \citet{hillier06} showed that the spectrum of \ec\ could be modelled only if the star and its wind suffered a visual extinction ($\approx 7$ mags at V) that was much larger than could be attributed to the interstellar medium ($< 2$ mags). However, the models of \citet{hillier06} always underestimate the observed EWs of some emission lines such as H$\alpha$ and H$\beta$. The inability to match these lines can be attributed to the presence of an occulter that reduces the continuum flux more than the line flux.}

{The presence of the coronagraph explains the unusual spectra taken in the 1980s and 1990s using ground-based spectrographs. The spectra exhibited broad hydrogen and permitted iron lines characteristic of a typical P\,Cygni star with a terminal velocity of $\sim$\,500\,km\,s$^{-1}$ \citep[e.g.][]{zanella1984,hillier92}. However, the spectrum also showed narrow forbidden lines of Fe superimposed on broad profiles. The narrow forbidden lines arise in partially ionised structures including the three Weigelt knots  \citep{weigelt86,davidson95}, while the broad components arise in the very extended wind \citep{hillier2001}. Normally the wind components would be too faint to be seen. However, the coronagraph systematically weakens the stellar continuum, but less so the broad components from the very extended wind. As the coronagraph vanishes, the stellar continuum emission increases and hence the extended broadened wind lines and the narrow forbidden lines, which have constant flux, become less prominent \citep{Mehner2010,davidson05}.}

{\citet{mehner12} made detailed monitoring of a set of spectral lines in the period 1998-2011, in direct light to the star and also at the FOS4 position \citep{davidson95} near the SE  pole of the Homunculus. In addition to confirming the weakening of the wind lines of the central star, that work showed that the broad-line absolute EWs formed in the primary's wind decreased at a faster pace than those observed in reflected light at FOS4. Part of the present work is focused on the equivalent width (EW) evolution of the three representative lines of the stellar wind (\ha, \hd and \ion{Fe}{ii} 4585{\AA}) studied by \citet{mehner12},  extending the data set to the time interval 1990-2021 (see Figure\,\ref{Ha-Hd-long}). The observational results of \citet{mehner12}, made during a shorter time interval, were confirmed, although the physical mechanisms suggested to explain the observations differ from those presented here.} 

{\citet{mehner12} suggested that the secular evolution of the spectra is due to variations intrinsic to the primary star that is still recovering from the 1840s Great Eruption. They suggest that the mass-loss rate is decreasing, making the wind more transparent. However, since the evolution of the spectral lines from the star and the corresponding ones at FOS4 are different, \citet{mehner12} postulate that the stellar wind at the equator is evolving differently from that at the pole. While the P~Cygni absorption component can be strongly dependent on viewing angle, the emission component in a moderately aspherical wind will vary much less, simply because one is averaging the emission over a large volume. Thus, it is unlikely that a simple latitudinal variation in mass-loss rate could explain the observed variations. Further, this model, by itself, does not explain the large changes in the apparent brightness of \ec.}

Recent ALMA observations have also provided direct confirmation of the coronograph at sub-arcsecond resolutions ($<$\,0\farcs5  $\sim$\,1100\,au). Although ALMA mapped molecular absorption lines,  they should trace the warm/cold dust. \citet{bordiu19} found CO ($-$9\,\kms), HCN (Hydrogen cyanide) and H$^{13}$CN ($-$60\,\kms) in absorption in front of the star at $T$\,$\sim$\,500\,K. \citet{morris20} found an absorption spot they attributed to CH$_3$OH (\citep{bordiu19}) and another of $^{12}$CO at radial velocity $+$3\,$\pm$\,5\,\kms in front of the star at $T$\,$\sim$\,100-110\,K. Different velocities and temperatures suggest that the absorbing material (inside the occulter) are spread at different radii in front of the star. These molecular absorption lines should be fading, but only future visits by ALMA can test this hypothesis.

In this work, we present new data and analyse spectroscopic evidence for the existence of a natural coronagraph in front of \ec.
The paper is organised as follows:
Sect.\,\ref{sectionobservations} presents the data; Sect.\,\ref{sectionresults} presents our results; in Sect.\,\ref{sectiondiscussion} we discuss these results; Sect.\,\ref{sectionconclusions} presents our conclusions.

Through this paper we use the binary ephemerides reported by \citet{teodoro16}: period\,$=$\,2022.7d, phase zero\,$=$\,HJD\,2456874.4. Following \citet{grant20} the {actual} periastron passage occurs four days earlier than this definition of phase zero.
The numbering of the orbital cycles was defined by \citet{groh04}.

\section{Observations}
\label{sectionobservations}

 We have collected a large number of medium/high-resolution spectra over the past 30 years from different observatories, by different observers and reported in a number of papers. 
 
 The reader should bear in mind that we do not aim to present a general description of the spectral features in \ec\ -- this has been done in earlier works.  We selected the data by focusing on a specific question: Is there any strong support for the existence of a natural coronagraph in \ec? Consequently, we do not explore spectral features and particular modeling results that are interesting but a distraction from our goal.

 Homogeneity of the data is a major concern in a large data set. Also of major concern when measuring EWs and line fluxes is the definition of the stellar continuum. This is particularly difficult in \ec\ since the lines are relatively broad, frequently display extended wings, and are often blended.
 \'Echelle observations can exacerbate the problem further because of the difficulty in joining orders. Signal-to-noise ratio (S/N) and spectral resolving power (R) are important for measuring line profiles, less so for measuring EWs. For measuring EWs, the wavelength integration limits are important, although subject to a somewhat arbitrary choice, due to blends with neighbour lines. This can be controlled by always using the same limits. The blended lines are much fainter and insert a low-level bias in the time series.  The seeing also is not too critical for the central star in \ec, since most of the emitting structures are confined to the central 0\farcs5, and the nebular contribution is far smaller.

 As a quality check on measurements, we have compared slit spectra taken at 1\arcsec-1\farcs5 seeing with FEROS fiber-fed (diameter 2\farcs7) and found no important line profile differences. We made extractions of the stellar spectrum in a typical UVES spectrum (seeing fwhm\,$\sim$\,0\farcs8) using apertures starting from 0\farcs5 (2\,pixels) up to 2\arcsec (8\,pixels) and found no significant differences in the normalised spectra. We then defined 1\farcs0 as the length along the slit to extract the UVES spectra.
 In order to check our procedure, we compared our measurements with those published in \citet{mehner12} and found reasonable agreement. Spectra from the star taken with different ground-based spectrographs coeval with UVES observations also showed good agreement without applying any adjustment shifts in EWs.

  Good image quality is of major importance for the nebular emission, as is the scattered light internally to the spectrograph. For this task, the UVES data set taken at the FOS4 position is crucial.

 %HST/STIS spectra resolve the structures around the central star, which demands special care in extracting the spectra to be compared with ground-based observations. Therefore, for the sake of homogeneity we did not use them, also because they do not exhibit a dense time sampling or long-term monitoring, in contrast with the ground-based data.
 HST/STIS spectra resolve the structures around the central star, which demands special care in extracting the spectra to be compared with ground-based observations. For the sake of homogeneity, since they do not exhibit a dense time sampling or long-term monitoring as the ground-based data, we did not use the HST data.

 \begin{figure*}
 \centering
 {\includegraphics[width=\linewidth, angle=0,scale=0.9,viewport=10bp 0bp 590bp 550bp, clip]{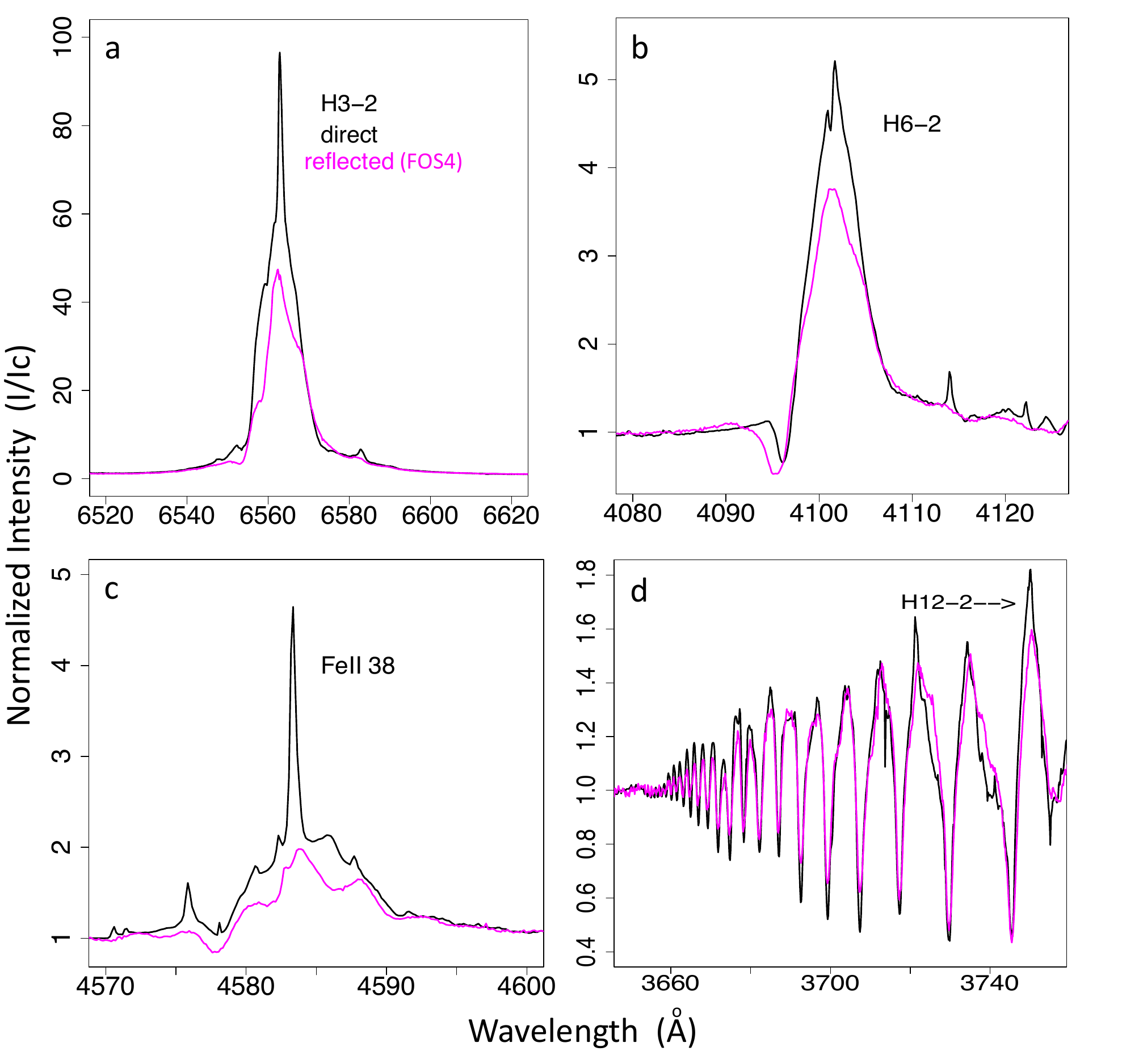}}
 \caption{{ Profiles of representative primary wind lines in direct (black) and reflected (magenta) views}. Spectra taken at high excitation phase ($\phi\,=\,$11.52, May/June 2006) with UVES. Panel a) \ha; b) \hd; c) \ion{Fe}{ii}\,$\lambda$4585{\AA}; d) High members of the Balmer series.  Radial velocity shifts have been applied in order to match the centroid positions of the emission lines in the reflected spectrum with those in direct light.}
\label{directxreflected}
\end{figure*}
 
 For the purpose of the present work, the main sources of data are described below.
 
\begin{itemize}
  \item {\it UVES direct and reflected light spectra:}
We selected high R and S/N spectra observed at VLT/UVES/ESO-Chile with $R$\,$\sim$\,100,000. The data were retrieved from the {\tt \ec Treasury Program}\footnote{\url{http://etacar.umn.edu/}} archive and were used to study the difference between direct and reflected-light spectra. We extracted UVES 1D spectra by tracing the 2D spectra within $\pm$2 pixels (1\arcsec) from the slit center. The slit width was 0\farcs4. We used coeval stellar and FOS4 spectra (when available) with the same parameters to extract the nebular spectra. Most of the UVES spectra at FOS4 \citep{davidson95, mehner12} in the period 1999.97-2009.50 were centered at $\sim$\,2\farcs84\,E and $\sim$\,$-$2\farcs36\,S of the central star (3\farcs7\,SE). Unfortunately, after 2013 this position was missed by the telescope pointing and the slit center was not in coincidence with FOS4. As shown in \cite{mehner12}, line profiles change from point to point. Although extracted the few nebular spectra in 2013-2015 at $>$\,0\farcs5 off the slit center, we did not use these in the formal analysis. EWs from FOS4 spectra are listed in Table\,\ref{table-uves-fos4}. 

A pair of spectra in direct and reflected light was taken in  May/June 2006, separated by 18 days, both during the high-excitation phase - when the cyclical variability arising from the 5.538\,yr period was at a minimum - to extract the representative line profiles (they are shown in Fig.\,\ref{directxreflected}). The observations of the direct spectrum were obtained on HJD\,=\,2453894.5, and for the reflected spectrum at the FOS4 position on HJD\,=\,2453866.5. Although not simultaneous with the reflected spectra, the inspection of \ha spectra taken in direct light at several observatories in the time frame 2005.5-2006.5 confirmed that \ha-EW was stable at better than 5\%.

\item {\it Optical spectra taken with the FEROS fiber-fed spectrograph ($R$\,$\sim$\,48,000) at the 0.52 and 2.2-m telescopes at La Silla Observatory (ESO-Ch):}
The spectra were obtained in the time-frame 1992-1999 \citep[][and references therein]{damineli00} and a few others in subsequent years. In these spectra, we measured EWs for \ha, \hd and \ion{Fe}{ii} 4585{\AA}. The fiber diameter projected onto the sky was 2\farcs7.

 \item {\it Optical spectra taken with the CTIO 1.5m telescope} with two different spectrographs:  CHIRON/CTIO ($R$\,$\sim$\,90,000) in the period 2009-2021 and the previous version of this spectrograph, a fiber-fed echelle with $R$\,$\sim$\,40,000; as described by \citet{richardson10, richardson15, richardson16, mehner12, teodoro16}. 
 
 \item {\it The SOAR/Goodman spectrograph} was used in several runs from 2008 to 2014 in the blue part of the spectrum, with $R$\,$\sim$\,3500 \citep{teodoro16}.

 \item {\it Optical spectra collected with NRES/LCOGT at South Africa and CTIO} with $R$\,$\sim$\,48,000. The fiber diameter projected onto the sky was 2\farcs8 (Navarete \textit{et al.}, in prep.). The 2019.9 LCO spectrum was used to compare with the 1999.1 FEROS spectrum listed in columns 5 and 6 of Table\,\ref{cmfgen1}. 

\item {\it Optical spectra were taken at the Coud\'e focus of the 1.6\,m telescope at the Pico dos Dias Observatory (OPD/LNA-Br)} 
In the optical range a variety of CCDs were used, covering $\sim$\,500 to $\sim$\,1000\AA\ at each grating angle. The response of the system as a function of wavelength is very flat and a simple linear fit can be used to rectify the stellar continuum in each spectral range, in contrast to the blaze function of the Echelle spectrographs. The room temperature of the spectrograph is very stable, resulting in accurate radial velocities. In the period 1989-2005 we used thick CCDs,  delivering  R$\sim$6,000. After the year 2005 we used 2048\,$\times$\,2048 thinned CCDs delivering R$\sim$ 12,000.  The typical slit width was $\sim$\,1\farcs5 and the seeing varied from  $\sim$\,1\farcs0 to $\sim$\,3\farcs0, although a number of observations were done at higher airmasses, which implies somewhat worse seeing.

 \item {\it NIR spectra were taken at the Coud\'e focus of the 1.6\,m telescope at the Pico Dos Dias Observatory (OPD/LNA-Br)}
to survey the \ion{He}{i} 10830{\AA} line since 1989. Before 2001 we used thick CCDs delivering  $R$\,$\sim$\,12,000, covering the spectral range $\lambda$10700-11000\,\AA. Although having efficiency lower than 1\%, these arrays did not exhibit noticeable fringes like the thinned CCDs, which were used after 2005.  The thinned CCDs delivered $R$\,$\sim$\,22,000 and covered approximately the same spectral range as the thick ones.  In the period 2001-2004 a HgCdTe infrared array was used, delivering $R$\,$\sim$\,7,000 \citep{groh07}, covering the spectral range $\lambda$10450-11000\,\AA.

 \item {\it A NIR spectrum taken at CRIRES/VLT/ESO-Ch} on 2009 April 03 and reported by \citet{groh07}  was used to explore the spectral absorption feature  at $\lambda$\,10792\AA. The resolving power was $R$\,$=$\,90,000 and spatial resolution 0\farcs33). 

 \end{itemize}
  
\section{Results}
\label{sectionresults}

\subsection{Spatially resolved line strengths}
\label{spatial}

We checked the finding by \citet{hillier92} who reported that spectral lines scattered by the Homunculus suffer less extinction than those observed in direct light towards the central star. UVES observations were carried out with sub-arcsec spatial resolution, not requiring complicated correction for the light scattered inside the spectrograph as in the previous work.  Figure\,\ref{directxreflected} shows representative lines we used.

A basic characteristic of the wind lines is that the higher the upper energy level, the closer the line originates to the primary's stellar core. Lines of lower excitation arise from increasingly larger regions of the primary’s wind (which dominates the wind emission), while higher excitation wind lines originate in the inner wind near the star. Thus, the comparison of continuum-normalised wind line-profiles of different excitation towards the star and reflected by the Homunculus can be used to map the spatial extent of the wind.

Panels a through c in Fig.\,\ref{directxreflected} display Balmer lines by increasing upper level. The similarity between the direct and reflected light improves from lower-level transitions (e.g. \ha and \hd, Fig.\,\ref{directxreflected}.a and Fig.\,\ref{directxreflected}.b) towards upper-level transitions (Fig.\,\ref{directxreflected}.c). We use here model parameters for the central star using CMFGEN, quoted in Sect.\,\ref{cmfgenparameters}. The \ha line (Fig.\,\ref{directxreflected}.a)  forms up to  large radii in the stellar wind ($r$\,$\sim$\,100\,R$_\ast$\,$=$\,12,000\rs\,$=$\,56\,au\,$=$\,24\,mas). Therefore its wind component is mostly insensitive to LOS absorption, contrary to the continuum flux, which increases towards the star.  Thus the apparent decrease in the continuum-normalised strength of the \ha line in the reflected spectrum (Fig.\,\ref{directxreflected}.a)  shows that the continuum suffers higher extinction in the LOS than in other directions. On the other hand, the \hd line arises from a smaller region around the star and therefore is sensitive to the same small-scale LOS absorption that also affects the stellar continuum.  Thus the temporal continuum-normalised  \hd line  in the direct spectrum is more similar to that reflected at FOS4 than the case of \ha\ (Fig.\,\ref{directxreflected}.a).

 In direct light, the upper-level transitions of the Balmer series (Fig.\,\ref{directxreflected}.d) are very similar to those observed in the reflected spectrum, which indicates that their formation region is entirely obscured by the coronagraph. The transition H\,12$-$2  at the right of Fig.\,\ref{directxreflected}.d is formed at $r$\,$\sim$\,10\,R$_\ast$, which poses a minimum radius for the coronagraph, arbitrarily assuming it to be circular. Figure\,\ref{directxreflected}.c shows the line profiles of permitted and forbidden singly-ionised iron around $\lambda$4585\AA. They are typically formed at $r$\,$\sim$\,200\,R$_\ast$, and suffer less extinction by the coronagraph than the stellar continuum, similar to   \ha. An extreme example of the contrast effect is the fact that the \ion{He}{ii} 4686{\AA} EW is repeatable from cycle to cycle, as reported by \citet{navarete20} and by \citet{teodoro16}. At periastron, the bulk of this Helium line is formed in a small region around the wind-wind collision (WWC) apex, unresolved in HST/STIS spectra.  Like the upper-level transitions of the Balmer series, its characteristic disk is smaller than the coronagraph, since  \ion{He}{ii} 4686{\AA} EW does not show the varying contrast effect between the stellar continuum and line emission. The \ion{He}{i} 7065{\AA} line displayed in panel e behaves differently from the others we have described. Its P Cygni absorption is stronger in direct than in reflected light because it has a contribution from the WWC walls \citep{richardson16}.

\begin{figure}
 \centering
 {\includegraphics[height=0.9\linewidth, angle=-90,viewport=40bp 120bp 915bp 460bp, clip]{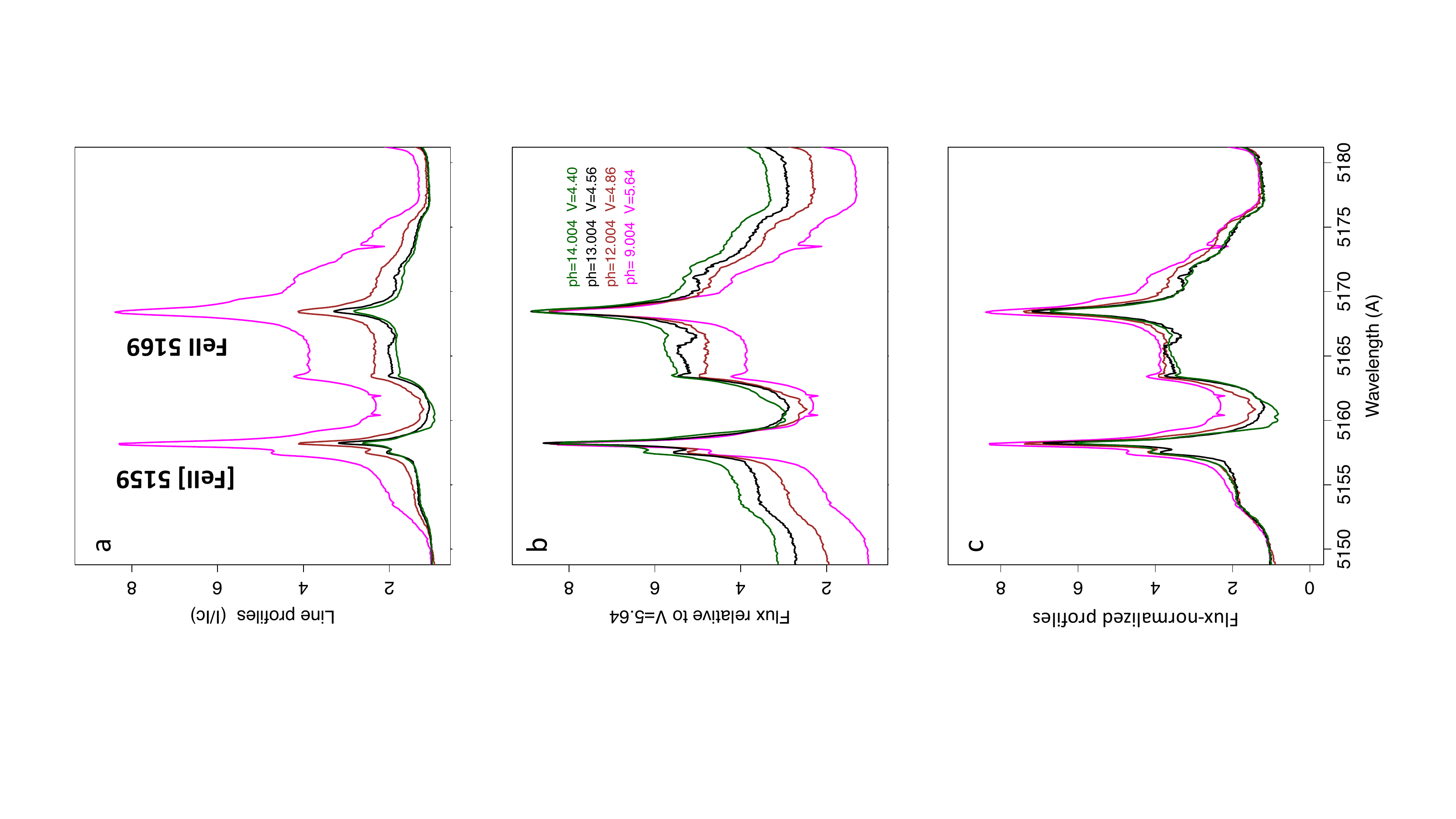}}
 \caption{{ Variation of \ion{Fe}{ii} $\lambda$5169 and [\ion{Fe}{ii}] $\lambda$5159 narrow and broad components at the same phase for four orbital cycles.} $\phi\,=\,$9.004 corresponds to 1992\,June\,20.
 a) Line profiles normalised to the stellar continuum showing the long-term decreasing strength caused by the secular brightening of the stellar continuum.
 b)  Line profile fluxes relative to the $\phi\,=\,$9.004 brightness ($V$\,=\,5.64\,mag). The narrow line components remained constant because they are formed in the Weigelt knots, outside the coronagraph.
 c) Flux-normalised line profiles. The broad line components, formed in the wind, are little affected by the brightening. 
 }
\label{FeII}
\end{figure}

The strong, narrow emission lines superimposed on the direct stellar spectra are produced by the Weigelt knots and other fainter dense structures spatially resolved but in close vicinity to \ec \citep{davidson95, Gull16}. They also appear in the spectra scattered by dust in the Homunculus, but are greatly attenuated and significantly broadened as they are intrinsically very faint. Their greatly decreased contribution in reflected light is another indication that the stellar continuum along our LOS to the primary core is largely suppressed.

Some features shown in these spectral lines may not be related to the depression of the stellar continuum,  and will not be further discussed. For example, the distortion on the line profiles, like the P\,Cygni absorption and the shoulder in the red arm of \ha $\lambda$6563\,{\AA} and \ion{Fe}{ii} lines, is likely due to a latitude-dependent stellar wind \citep{Smith_2003},  or the effects of the wind cavity \citep{groh12}, and inherent broadening due to velocity dispersion of the scattering dust.

\subsection{Cycle-to-cycle weakening of the EW of the primary's wind lines}
\label{temporal}

We demonstrate in Fig.\,\ref{FeII} how the continuum level has changed relative to line profiles with two typical, singly-ionised iron lines  $\lambda\lambda$\,5159 and 5169\AA. Note that the first transition is permitted and the second is forbidden. They have been observed at nearly the same phase over four orbital cycles. This figure shows line profiles observed at 9 days after phase zero, with phase $\phi\,=\,$9.004 corresponding to 1992\,June\,20. Figure\,\ref{FeII}.a shows the rectified spectra for a region containing both a permitted and a forbidden transition. Both lines exhibit a broad component formed in the wind, and a narrow component that is formed in the BCD Weigelt knots located at $>$\,0\farcs25 to the NE of the stellar core. As shown in this plot, the line's absolute EWs have decreased with time.

Figure\,\ref{FeII}.b shows these two lines' flux corrected using the $V$-band coeval magnitudes of the Homunculus (in the upper right) and scaled to unity at phase $\phi\,=\,$9.004 when the brightness was $V$\,=\,5.64\,mag.  The flux of the narrow-line components remained constant, confirming that the Weigelt knots remained at a constant brightness. This demonstrates the complementarity of ground-based observations to STIS/HST resolved spectra, as the space-based observations were the first to reveal the roughly constant brightness of the Weigelt knots \citep{hillier06, Gull09, Mehner2014}.

 Figure\,\ref{FeII}.c shows flux-normalised line profiles, which is the same as  Figure\,\ref{FeII}.b when it is normalised to unity at the stellar continuum. This plot shows that the broad line components formed in the wind are little affected by the brightening. The broad lines have changed only slightly with time and, in particular, there is no indication for a change in the speed of \ec's wind. Since the Weigelt knots were not affected by the changes in the continuum flux, the narrow line components in Figure\,\ref{FeII}.c are off-scale.
 
  The natural explanation for these features is that the major change with time is caused by decreasing extinction due to dissipation of dust and/or tangential motion of the obscuring material {covering} the stellar core and possibly the extended structure located to the SE, opposite the positions of Weigelt C and D relative to \ec.
  
\subsection{Differential weakening of Balmer lines for the central core}
\label{differential}

As we have extensively monitored the star, especially across periastron passages since  1992 (periastron\,9), we can compare the evolution of line profiles at the same binary phases. Spectra recorded at phase $\phi$\,$\approx$\,0.025 ($\sim$50 days after phase zero), have been taken when the secondary star is deeply embedded in the extended primary wind as it passes behind the primary star. Thus, we are able to probe the relatively undisturbed wind of the primary star for a brief time (days to a few weeks) across each periastron event. 

\begin{figure}
 \centering
 {\includegraphics[height=0.9\linewidth, angle=-90,viewport=30bp 0bp 950bp 550bp, clip]{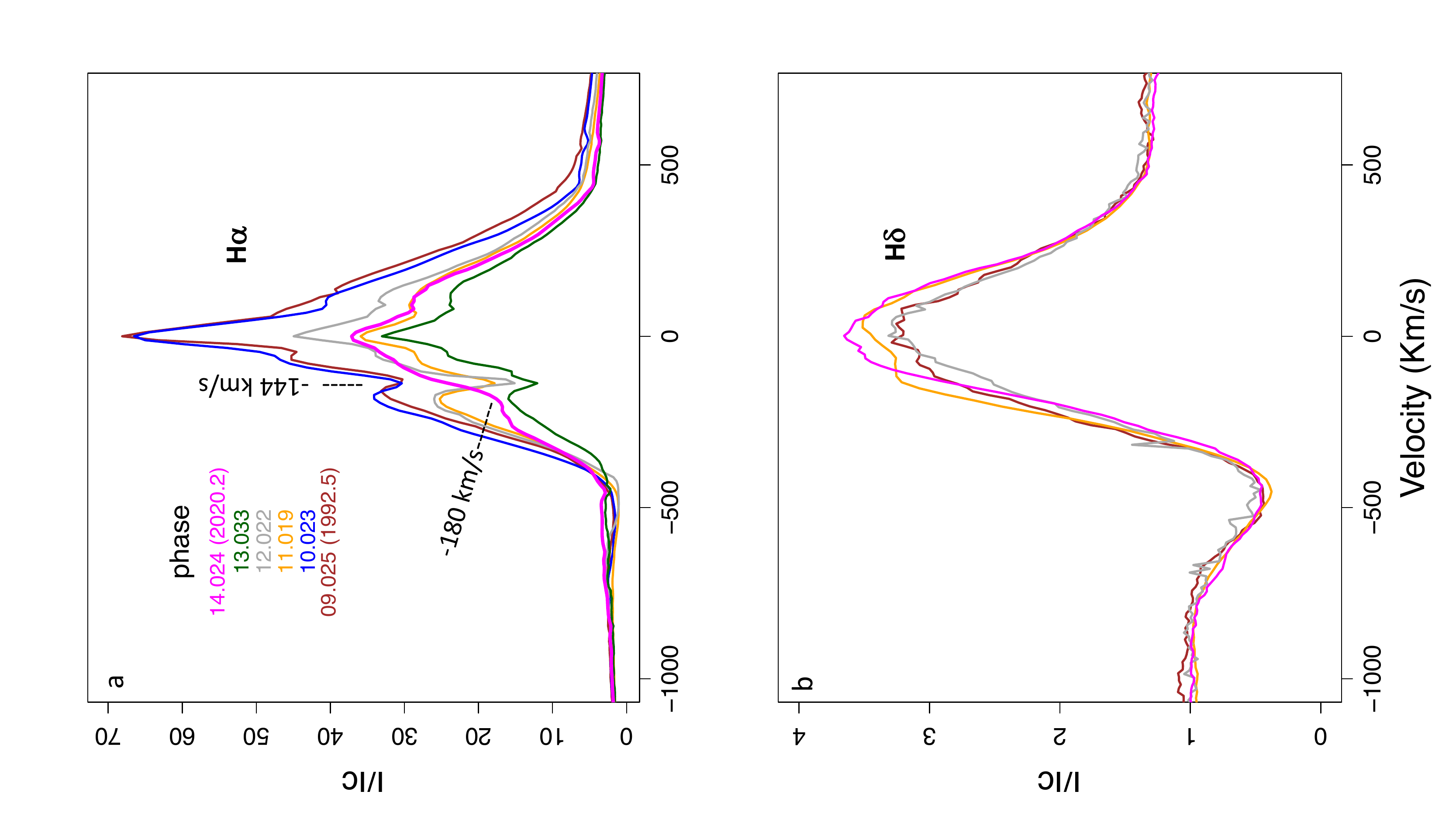}}
 \caption{{ Evolution of \ha and \hd at periastron. 
 Line profiles of  \ha (panel a) and \hd (panel b) during the last six orbital cycles at the same phase $\phi$\,$\approx$\,0.025 ($\approx$\,50 days after periastron). The \ha line peak decreased by $\approx$\,50\% while \hd remained nearly constant. The constancy of the \hd P\,Cygni absorption depth, terminal velocity and emission indicate a stable mass-loss rate. There are no \hd observations at this phase for cycles \#10 and \#13. The same colours in both panels correspond the same observing dates.}}
\label{hahdel}
\end{figure}

Figure\,\ref{hahdel}.a shows the evolution of the normalised \ha line-profile across six periastron events. The strongly continuum-emitting core has become much more visible, causing the absolute EW of \ha to decrease. By contrast the \hd line shows only minor changes in the emission profile (Fig.\,\ref{hahdel}.b), indicating that it is formed primarily behind the coronagraph. Furthermore, the blueshifted P\,Cygni absorption, which is produced along our LOS to the star, also has not changed. The constancy of the profile indicates that the terminal velocity, the mass loss, and the luminosity  (by inference) of the star, have not changed very much over the last 30 years.

\subsection{Evolution  of line strengths in direct and reflected light}
\label{direct-reflected}
\begin{figure}
 \centering
 {\includegraphics[height=0.95\linewidth, angle=-90,viewport=20bp 100bp 770bp 450bp, clip]{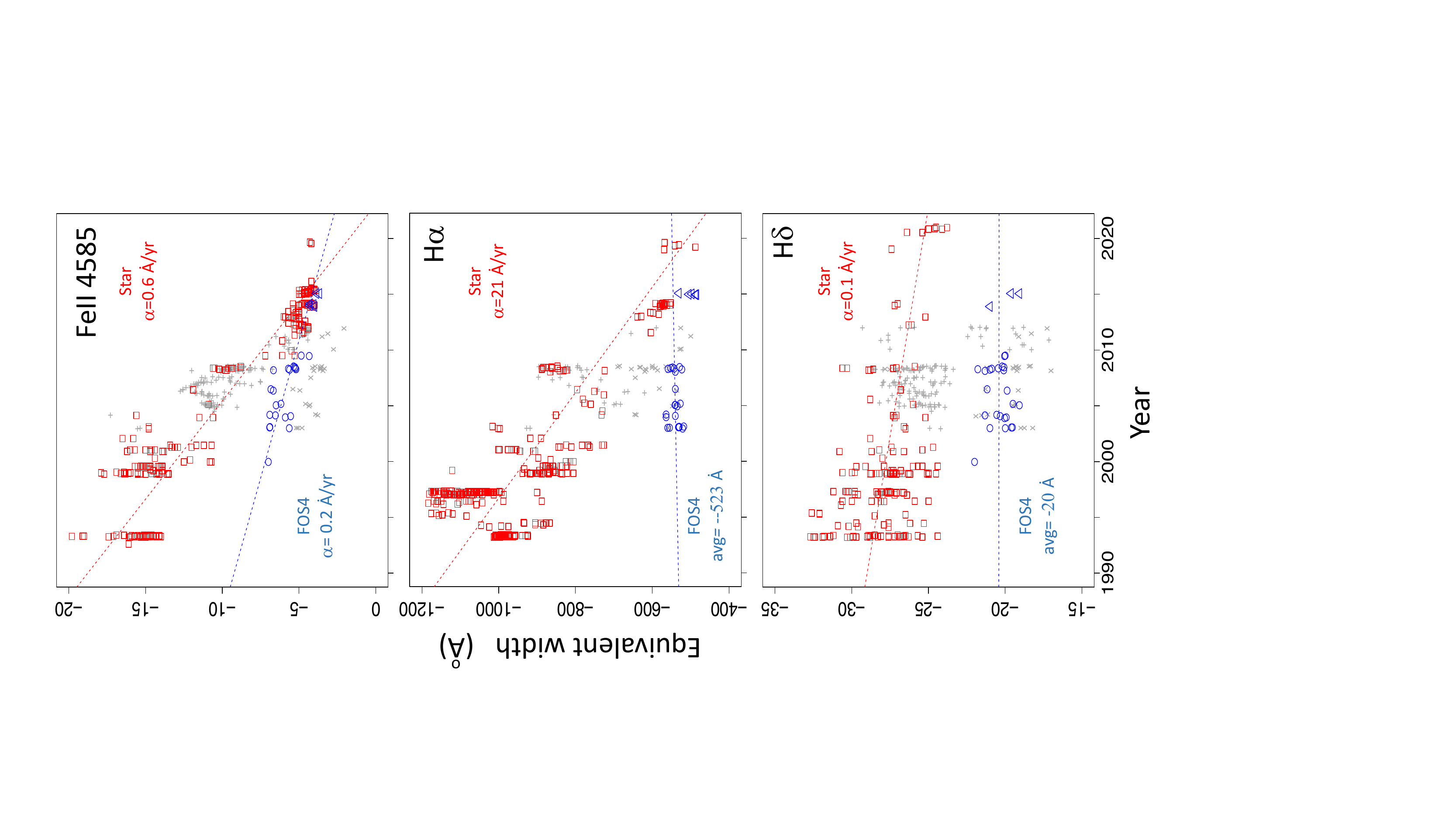}}
\caption{{ Cycle-to-cycle evolution of the equivalent width of representative primary-star wind lines in direct (red) and reflected view at FOS4 (blue). Top panel) \ion{Fe}{ii} 4585\,\AA\ has the steepest gradient of growing fainter in both views. Middle panel) \ha direct shows an intermediate decrease rate of the absolute EW and  constant level  $\sim$\,$-$523\,\AA at FOS4. Bottom panel) \hd in direct view shows a very gentle slope and at FOS4 the line is almost constant at EW\,$\sim$\,$-$20\,\AA. Gray points indicate measurements reported by \citet{mehner12}, with $+$ indicating direct view and $\times$ reflected at FOS4. Measurements in the range of phases 0.93-1.07 were omitted in order to highlight cycle-to-cycle variations. Blue triangles indicate EWs for recent years when the telescope pointing was too far from FOS4; these were not used in the linear regression fits.}}
\label{Ha-Hd-long}
\end{figure}

 {

{
Figure\,\ref{Ha-Hd-long} shows the time evolution of three representative lines of the primary's wind from 1990 to 2021: \ion{Fe}{ii} 4585\,\AA\ in the top panel, \ha in the middle and \hd at the bottom.  The EW absolute value\footnote{the absolute value is used because the classical definition of EW is negative for emission lines} has decreased in some cases and remained constant in others. To aid in the interpretation of the variations
we fitted the data (using least squares) to a simple linear
function of the form $EW=a(t-2000)+b$, and the coefficients of the
fits are provided in Table \ref{table-rate}.

Tables \ref{table-direct-ha}, \ref{table-direct-fe2} and \ref{table-direct-hd} show all measurements for the star from all observatories.  EWs were measured for the entire set of spectra using the same parameters as those defined by \citet{mehner12} for the continuum normalisation and EW integration. The integration range was $\lambda$\,6520-6620\,\AA\ for \ha, $\lambda$\,4085-4115 for \hd\ and $\lambda$\,4570-4600\,\AA\ for the \ion{Fe}{ii}\,$\lambda$4585\,{\AA} line. 
}

To place lines of different strengths on a more equal footing we
use the EW ratio between the years 1995 and 2020 (R95/20 in column 4 of Table\,\ref{table-rate}) to compare variations. The
\ion{Fe}{ii} 4585\,\AA\ absolute EW decreases at a pace $\sim$\,6 times faster than \ha and $\sim$\,11 times faster than \hd. The far smaller changes in \hd EW (which is formed close to the star) compared to that of \ha or \ion{Fe}{ii} 4585\,\AA\ (formed far in the wind) indicates that \hd is being emitted in a region that is mostly covered by the coronagraph.
It also suggests that across the continuum and \hd formation regions the occulter is fairly uniform. However, it does not need to be concentric to the star.
}

\setlength{\tabcolsep}{4.5pt}
\begin{table}	
	\caption{ Absolute EW rate of decrease compared to the formation radius for spectral lines in direct view  to the star}
	\label{table-rate}
	\begin{tabular}{lrrrr} 
    \hline				
Feature     &slope$^a$ & intercept$^a$  & R95/20$^b$~ & log(r/R*)$^c$ \\
Units           &\AA.yr$^{-1}$& \AA  & & - \\
\hline
\hd               &0.12&-27.5& 1.1 & 1.5 \\
 \ha              &21.13&-990.4& 1.9& 1.9 \\
\ion{Fe}{ii} 4585 &0.581&-12.9& 12.3& 2.3  \\
\hline
\end{tabular} \\
{{Notes:}\,$^a$\,~\,linear fit EW (\AA)\,=\,a\,.\,(year\,-\,2000)\,+\,b\\
$^b$\,~\,R95/20\,=\,EW(1995)\,/\,EW(2020) using the linear fit\\
$^c$\,~\,CMFGEN model described in Sect.\,\ref{cmfgenparameters}
}
\end{table}
\setlength{\tabcolsep}{6pt}

{
Table \ref{table-uves-fos4} contains EWs of the same three lines reflected at FOS4. At this position, the decreasing rate was $-$0.2\,\AA\,yr$^{-1}$ for the absolute EW of \ion{Fe}{ii} 4585\,\AA; the \ha line strength remained constant at $-$523\,\AA; and \hd also remained constant at $-$20\,\AA. The blue triangles at the right end of the FOS4 series were extracted from spectra centered at distances larger than 0\farcs5 from FOS4. Although they follow the series reasonably well, were not used in the calculations.

Changes in the \ha EW measured in direct light seem to have stalled during the last orbital cycle (after 2014.5), with a value approaching the EW of the Homunculus-reflected light.  This is expected when the coronagraph has completely { vanished}. But this is earlier than predicted by \cite{damineli19} and may be due to a temporary halt in the dissipation/shifting of the coronagraph. 
}

\subsection{The blue-displaced absorption (shell-like feature) in the \texorpdfstring{H$\alpha$}{H-alpha} line profile}
\label{nar}

\begin{figure}
 \centering
 {\includegraphics[width=\linewidth, angle=0, viewport=210bp 30bp 760bp 480bp, clip]{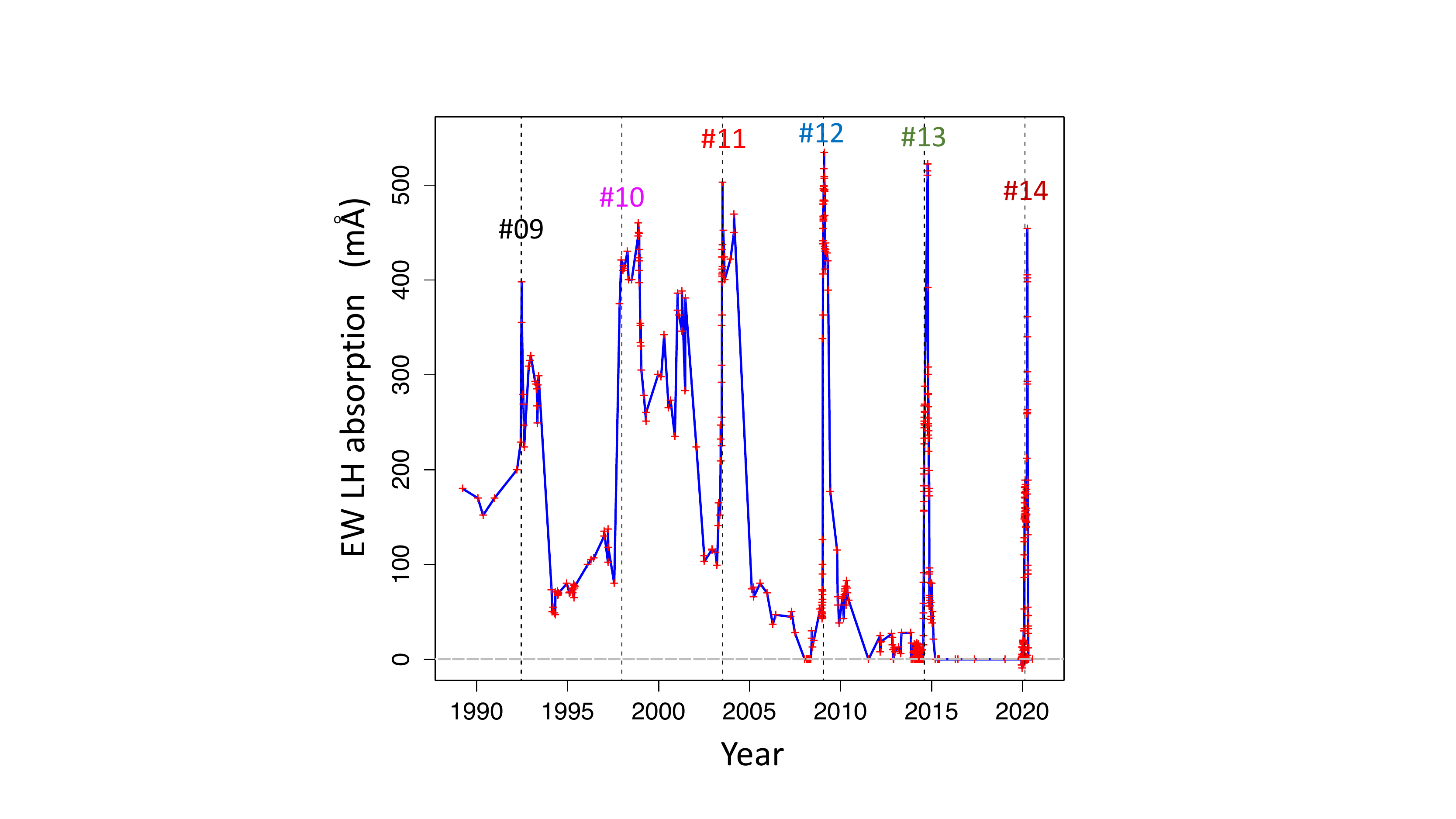}}
\caption{{ Historical EW time series of the blue displaced absorption on the \ha line profile - originated in the LH - showing three components: a) the periodic peak at periastron, b) occasional excursions to high intensity after periastron, and c) a relatively low-intensity component always present, but fading in the long term. Labels $\#$09 to $\#$14 indicate the orbital cycles. The EW was measured by direct integration over the \ha line profile, in the range $\lambda$\,6548-6556\,\AA.  }}
\label{Hanar-historic}
\end{figure}

The blue absorption feature in the \ha line profile (see Fig.\,\ref{hahdel}.a) is a signature of material in our LOS, which might be related to the coronagraph. It was first reported by \citet{ruiz1984} and detected periodically in all subsequent low excitation events that occur near inferior conjunction of the primary \citep{damineli1998, richardson10, richardson15}. It is not intrinsic to the primary wind; however, its velocity is consistent with that of the Little Homunculus (LH) nebula \citep{ishibashi03, Gull06}. When the system is outside periastron, the LH outside the LOS is fully ionised by the FUV emitted by the secondary star and the WWC and escaping through the shock hole. However, the ionised hydrogen recombines when the dense wind of the primary inhibits the escape of the photons from the secondary, but allows Lyman\,$\alpha$ to scatter in the circumstellar medium \citep{johansson05}. This occurs when there is a neutral or partly ionised intervening gas in our LOS, which produces a typical shell-like absorption, similar to that observed in some Be stars. The shell-like feature which has  RV\,$=$\,$-$144\,\kms,  FWHM\,$=$\,25\,\kms, was replaced by a broader component in cycle \#14, with  FWHM\,$=$\,82\,\kms\, centered at RV\,$=$\,$-$180\,\kms, which continues to be compatible with the LH velocity. 
This indicates that the primary's wind, or the region of the LH causing the  shell-absorption, suffered a change during the last periastron.

Figure\,\ref{Hanar-historic}  presents the EW historic time series of the shell-like feature associated with \ha. The dashed lines indicate phase zero of the periodic events. In addition to the periodic component, there are high-intensity excursions of the EW after phase zero of cycles \#09, \#10 and \#11. These mid-cycle peaks have duration of {four years} or less and might be due to dense structures passing across our LOS to the secondary star. They appear to have not been present in cycle \#13. The third feature in this time series is a low-level absorption component which was present outside periastron and has been decreasing in EW with time. It remained at EW\,$\sim$\,0 after the low excitation event \#13. 

The periodic peaks of the blue \ha absorption EWs (Figure\,\ref{Hanar-historic}) are not produced by the coronagraph, which varies on longer timescales. The occulter that causes the periodic peaks around periastron is the primary's wind, produced when the secondary star moves behind the primary near periastron. The shell-like absorption is also observed outside periastron and has a long-term behaviour. It is probably produced by other regions around the binary system. The long-term variable shell-like absorption is produced by the coronagraph. The complex behaviour of the shell-like absorption indicates that there are other occulters located in the circumstellar medium, located farther from the primary star but closer than the LH.

A similar narrow absorption line also exists in other directions, as reported by \citet{boumis98}. That direction could cross denser circumstellar material inside the LH that is sampled by our LOS and must be permanently neutral or partly ionised. This means that the UV radiation emitted through the WWC cavity never illuminates that direction.  That shell-like absorption is reflected on a small nebula at $>$2\arcmin to the NW of \ec, outside the Homunculus, and is seated at  RV\,$=$\,$-$80\,\kms on the \ha line profile. It is stronger than its sibling along our LOS and has not varied during 1985-1997. We took one spectrum with the Goodman/SOAR spectrograph in 2018 at the same spot as observed by \citet{boumis98}. Although our spectra have lower spectral resolution, no major changes were seen. This seems to indicate that the shell-like feature sampled in our LOS is part of a much larger structure.

\citet{pickett2021} reported a similar decrease of the  EW weakening of the  $-$144\,\kms narrow absorption component of Na\,D. Based on STIS/HST UV spectroscopy, Gull \textit{et al.} (in prep.) noted similar disappearance of narrow absorption lines. The HST/STIS spectra also show that the ionisation is increasing not only in our LOS, but also within the SW Homunculus lobe, changing dramatically to the NW of the stellar core. This indicates that the coronagraph may be extended, with its edge projected against the LH. The vanishing occulter modulating the FUV must be within the LH. The remaining occulting material in the coronagraph also affects recombination lines, but at larger distances, beyond the LH.

\subsection{A circumstellar DIB at  \texorpdfstring{$\lambda$10792\,{\AA}}{at 10792 Angstrom}?}
\label{DIB_id}

\begin{figure}
 \centering
 {\includegraphics[width=\linewidth, viewport=180bp 30bp 780bp 530bp, clip]{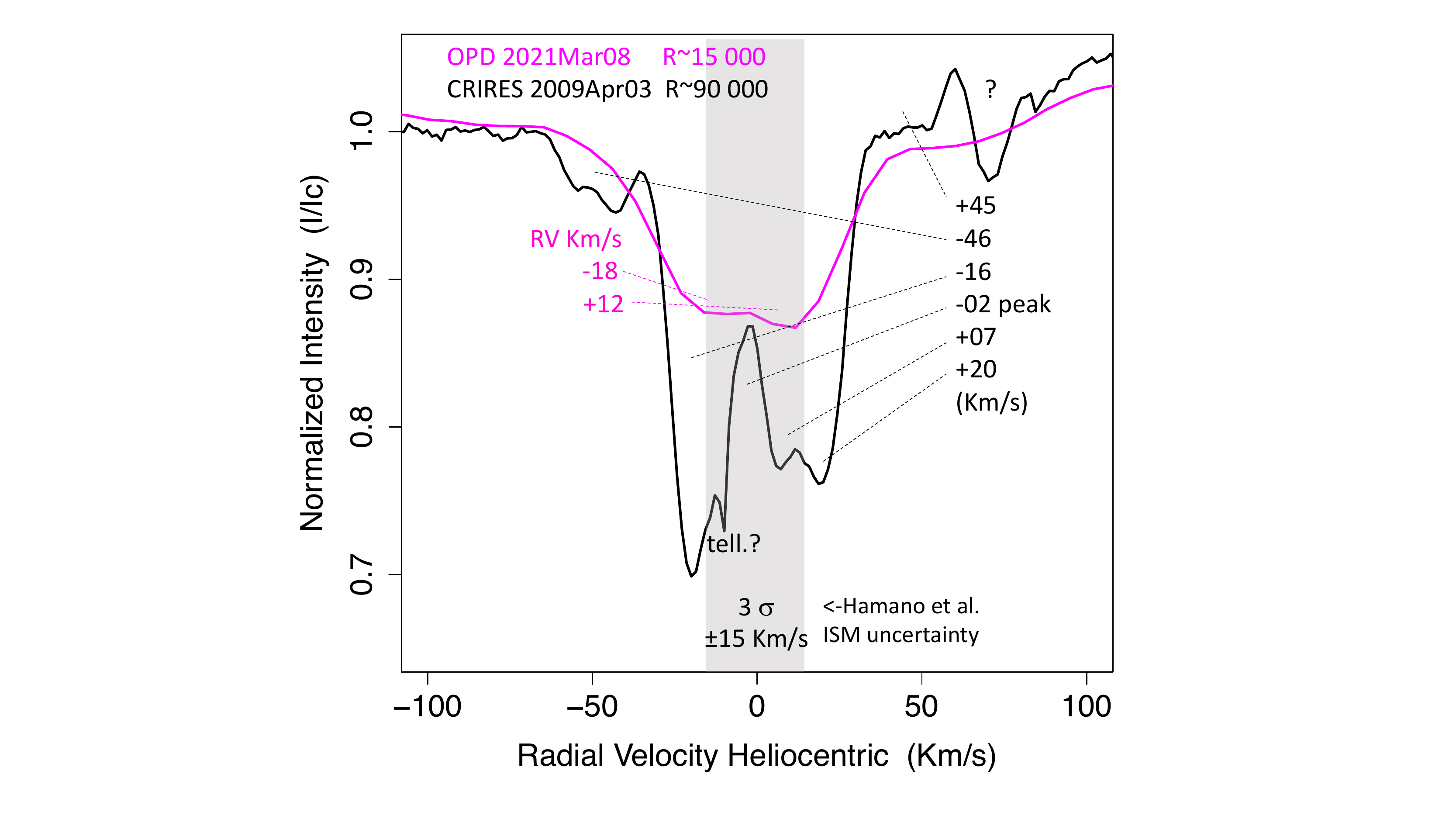}}
 \caption{{ Profile of the absorption line at $\lambda$\,10792\,{\AA}. The black line is from a CRIRES spectrum taken on 2009 April 03, showing at least four components blueshifted and redshifted between $\pm$\,45\,\kms from the central reversal. Zero velocity corresponds to the position measured by \citet{hamano15} for a diffuse interstellar band at the same position and the grey strip marks 3\,$\sigma$\,uncertainty in their measurements. The magenta line shows an observation at OPD/Coud\'e on 2021 March 08, with four times lower spectral resolution. It seems that the two main components of the absorption line changed the intensity ratio between 2009 and 2021.}
}
\label{DIB2-ident}
\end{figure}

\begin{figure}
 \centering
 {\includegraphics[width=\linewidth, viewport=10bp 30bp 615bp 520bp, clip]{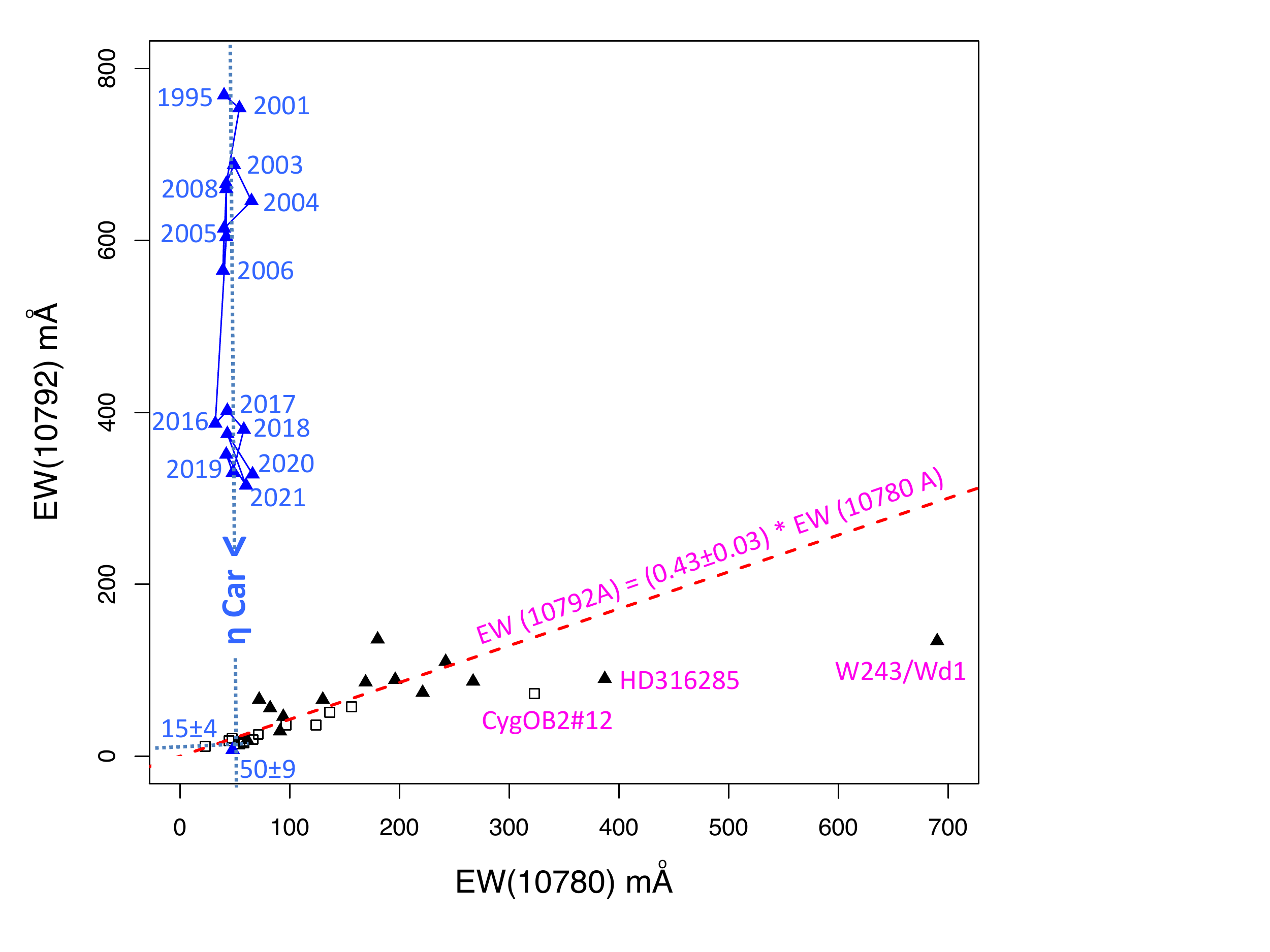}}
 \caption{{ Diffuse Interstellar Bands (DIBs) at $\lambda$10780\,{\AA} (DIB1) and $\lambda$10792\,{\AA} (DIB2) in a sample of stars. Squares are from \citet{hamano15} and triangles from this work. Points labelled in magenta indicate deviating objects from the linear relation between the two DIBs: EW(10792\,{\AA})\,$=$\,(0.43\,$\pm$\,0.03)\,EW(10780\,{\AA}). Blue triangles indicate a long-term fading of the $\lambda$10792\,{\AA} absorption in \ec\ (here called eta1.079}). }
\label{DIB2}
\end{figure}

{
We have explored the possibility that the isolated absorption feature at $\lambda$\,10792\,{\AA}, hereafter eta1.079, is a circumstellar counterpart of a diffuse interstellar band (DIB) \citep{diaz15}.  In addition to the agreement in wavelength position, its long-term strength decrease is well correlated with the decreasing extinction, and it has a gradient opposite to the P Cygni absorption strength of  \ion{He}{i} 10830{\AA}  and most of the resonant lines in the optical range.
}

{
This feature was reported in \ec by \citet{damineli93} who attributed it to a blue displaced absorption at RV\,$=$\,$-$1040\,\kms\, associated with the \ion{He}{i} 10830{\AA} line as also adopted by  \citet{groh10}. However, no similar feature at this velocity was present in the \ion{He}{i} 20587{\AA} ``sister'' line shown in that work and any other \ec spectral line reported {so far}.
}

{
 In order to examine its profile and wavelength, we plot a high-resolution spectrum reported by \citet{groh10}  in Figure\,\ref{DIB2-ident}, taken on 2009 April 03. Two other CRIRES spectra (from 2008 May 05 and 2009 Feb 09) reported in that work show a similar line profile close to $\lambda$\,10792\,{\AA} except for a central peak above the continuum, which might be a defect, not necessarily a real emission. The absorption feature at this wavelength shows a complex profile, with at least four components between $+$45 and $-$46\,\kms. There is a central reversal feature in almost exact coincidence with the $\lambda$10792 DIB reported by \citet{hamano15}.  The vertical gray strip in Fig.\,\ref{DIB2-ident} shows the 3\,$\sigma$ uncertainty around $\lambda$10792.15\,$\pm$\,0.15\,\AA\ reported by \citet{hamano15} for many directions in the Galaxy - hereafter named DIB2.

 A recent observation taken at the OPD/Coud\'e focus on 2021 March 08 with  R\,$=$\,15,000, four times lower than CRIRES, is also shown in Figure\,\ref{DIB2-ident}. The low amplitude reversal is at the same velocity as in CRIRES. The two main absorption components are shown at approximately the same velocity, but with interchanged depth ratio, in addition to this variation, the EW has been decreasing slowly over the long term, as described below. 
 }
 
 {
 Although the large positive velocity makes the identification of this feature as DIB2  unlikely, in order to gain more information, we measured the DIB2 on {many} normal and emission-line stars with spectra taken at the OPD/Coud\'e (Table\,\ref{table-DIBs}). Figure\,\ref{DIB2} shows our measurements of DIB2 compared with the closest one at $\lambda$10780\,\AA\ (hereafter DIB1). Figure\,\ref{DIB2} shows a very good linear correlation for most of the observations of ours and of \citet{hamano15}: EW(DIB2)/EW(DIB1)\,$=$\,0.43\,$\pm$\,0.03. These authors found a single discrepancy of the correlation for CygOB2\#12 and we show two additional exceptions for the LBVs HD\,316285  and W243 in the Westerlund 1 cluster. These measurements are based on sufficiently large EWs, making them quite reliable.  These three objects have much larger extinction and larger DIB1 EW, but have their DIB2 EW much fainter than expected. This indicates that DIB1 and DIB2 only mimic belonging to the same family of carrier molecules when measured in the solar vicinity but might have different behaviour in other parts of the Galaxy.
}

{
 Figure\,\ref{DIB2} includes measurements of eta1.079  since 1995 (Table\,\ref{table-eta1.079}). The feature eta1.079 is exceptionally strong compared to DIB1. When comparing its large EW with CygOB\#12, HD\,316285  and W243, which deviate in the opposite sense with respect to the linear fit, it seems unlikely that eta1.709 is produced by the same carriers as DIB1. However, there are only a few cases studied at very high reddening and we cannot reach a firm conclusion. {Despite the} relatively large errors in EW, the long-term behaviour of this feature in \ec\ is clear: its intensity is decreasing with time. On the other hand, the strength of DIB1 is consistent with constant EW. eta1.079 has been decreasing at a rate of $\sim$\,$-$10\,m\AA\,yr$^{-1}$ which evolves on the contrary sense to the P Cyg absorption in many spectral lines. For example, the P\,Cyg absorption of  \ion{He}{i} 10830{\AA} in the same period increased at a rate $\sim$\,$+$\,500\,m\AA\,yr$^{-1}$. If eta1.079 is not produced by DIB1 carrier molecules, it might be due to another mixture of molecules and does not seem to be an absorption associated with the primary wind.
}

\subsection{Correlated decrease of \texorpdfstring{$A_{\rm V}$}{V-band} extinction and the eta1.079 absorption within the coronagraph}
\label{subDIB2}

\begin{figure}
 \centering
 {\includegraphics[clip, width=\linewidth, viewport=100bp 20bp 620bp 470bp]{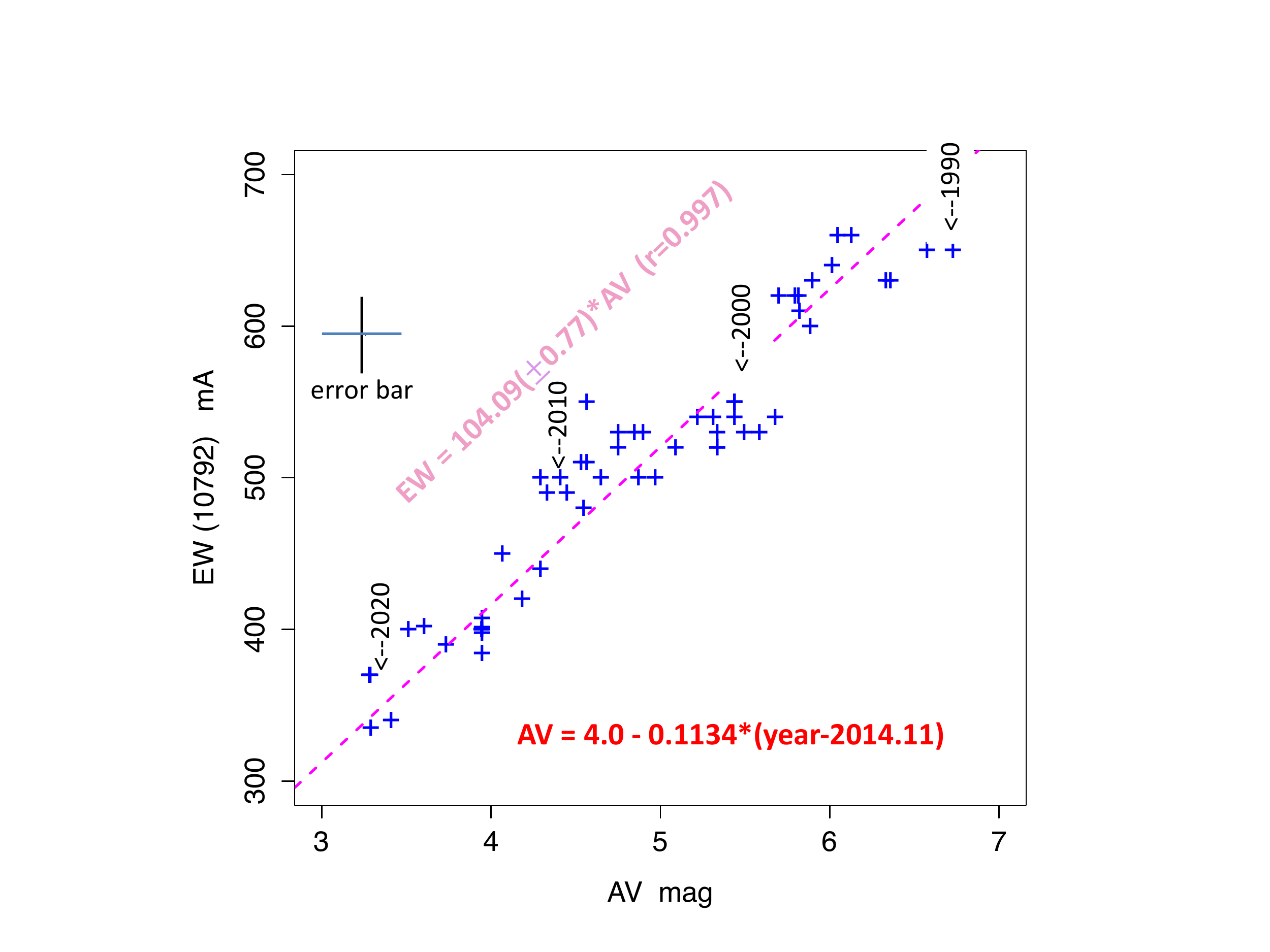}}
 \caption{{ Relation between eta1.079 ($\lambda$10792\,{\AA} absorption)  and $A_{\rm V}$}. The extinction towards the stellar core ($x$-axis) was reported by \citet{damineli19}. It is fading in the form $A_{\rm V}$(mag)\,=\,4.0\,$-$\,0.113\,(year\,$-$\,2014.11)  and is predicted to reach the ISM value $A_{\rm V}$\,$\sim$\,1.5-2.4\,mag in 2028-2036. The equivalent width of eta1.079 absorption ($y$-axis) is fading in tight correlation ($r$\,=\,0.997) with the visual extinction of the central star: EW(DIB2)\,(m\AA)\,=\,104.09\,$A_{\rm V}$. The typical error bar of the individual measurements is shown as a cross in the top-left of the plot (see Table\,\ref{table-eta1.079}).}
 \label{AVDIB2}
\end{figure}

{ 
\citet{damineli19} showed that the rate of decrease in extinction follows the equation $A_{\rm V}$\,(mag)\,=\,4.0\,$-$\,0.113\,(year\,$-$\,2014.11). If the excess absorption of eta1.079  in our LOS to the central star originates in the coronagraph, it would follow the extinction decrease. We used the above relation to derive $A_{\rm V}$ for the dates we have measured EW(eta1.079) and found a surprisingly tight correlation: EW(eta1.079)\,(m\AA)\,=\,104.09\,$A_{\rm V}$\,(mag), with r\,$>$\,99\% (dashed magenta line in Fig.\,\ref{AVDIB2}). Spectra taken in the time interval 1990-2021 are labelled for every decade in the plot to indicate how the changes vary with time.  Despite the relatively large errors in EW measurements (see Table\,\ref{table-eta1.079}), we see that the changing EW  of eta1.079  follows the decreasing extinction inside the coronagraph. }

The normal way to produce extinction is by dust, and so the correlation found in Fig.\,\ref{AVDIB2} might suggest that the eta1.079 carrier molecules are associated with dust grains or at least they are mixed inside the same region. The brightening in the $V$-band between orbital cycles 10.74 (2002.07) and 13.69 (2018.44) was 1.86\,mag as compared with 2.9\,mag at 1200\,{\AA} measured in STIS spectra (Damineli \textit{et al.}, in prep.).  If the brightening was due to just decreased extinction, the increase in brightening at 1200\,{\AA} should have been 8 magnitudes for a canonical reddening law ($R_{\rm V}$\,=\,3.1). This indicates that the extinction of the occulter was mostly grey, which requires very large dust grains or electron scattering in an ionised gas. Large dust grains are supported by dust model results of \citet{morris17}, in particular iron grains in the range of few to tens of microns. Such large grains would condense close to the central source at temperatures of a few hundred to $\>$\,800\,K and are not in thermal equilibrium. Such a high temperature in principle is against the identification of eta1.079 as the same carrier molecule as DIB2, which is formed in the very cold ISM. 

\subsection{Spatial structures inside the coronagraph}
\label{structures}

\begin{figure}
 \centering
  {\includegraphics[clip, width=\linewidth, viewport=220bp 20bp 820bp 530bp]{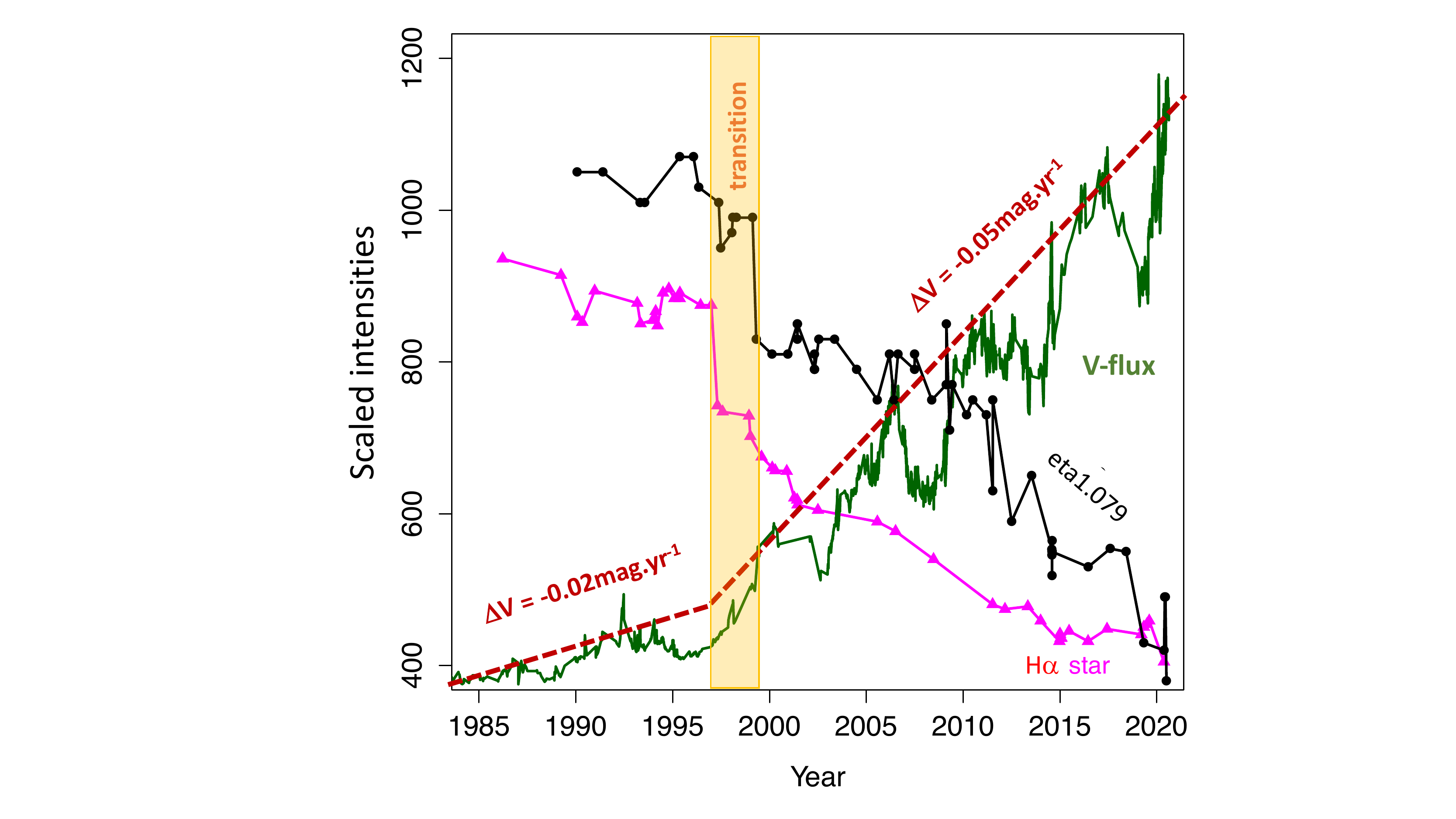}}
 \caption{{ Structures in the coronagraph.}
  The long-term brightening of the entire object (star + Homunculus)  due to decreasing extinction towards the stellar core is shown by the green line. Also shown are the EW of eta1.079 ({\it black line}) and the absolute EW of \ha ({\it magenta line}), both measured in the star. A strong inverse correlation between the EWs of eta1.079 and \ha with the $V$-band flux is observed.  The  $V$-band light curve (green line) is based on data taken from \citet{damineli19} and transformed to relative flux. }
 \label{figstructures}
\end{figure}

Figure\,\ref{figstructures} compares \ha\ and eta1.079 EWs plus $V$-band flux  variability from 1985 through 2020. The $y$-axis was scaled to represent the measured absolute \ha EW in \AA. Both the eta1.079 EW  and the $V$-band flux values were multiplied by constants to fit within the plot. The fading of the  eta1.079 EW (black line) occurs in concert with the  \ha EW (red line). The $V$-band flux of the whole object (stellar core + Homunculus, green line) increases with time. A well-known change in the brightening rate occurred in 1997-99. It seems that there was a coeval fading of \ha and eta1.079, labelled as ``transition'' in Fig.\,\ref{structures}. If this transition was as sharp as it appears to be, it would not be consistent with systematic coronagraph dissipation but implies a clumpy portion of the coronagraph moving out of our LOS.  

\section{Discussion}
\label{sectiondiscussion}

\subsection{A model for the central star}\label{cmfgenparameters}
    
To facilitate the discussion we present a model for the primary star, assuming a distance of 2.3\,kpc to the object \citep{AH93etashape, meaburn93,DSG01_shape,shull21}. 

Following previous works \citep{hillier2001,hillier06} we chose as basic parameters of the model:\\
$R_{\hbox{\rm core}}\,=\,120$\,\rs\\
$\dot{M}\,=\,8.0 \times 10^{-4}$\,\Msunyr\\
$L=5.0 \times 10^6$\,\ls\\  
$V_\infty=420$\,\kms\\

These are similar, but not identical
to earlier values adopted by \cite{hillier2001}:\\
$R_{\hbox{\rm core}}=60$\,\rs\\
$\dot{M}$\,=\,1.0\,$\times$\,10$^{-3}$\,\Msunyr\\
$L$\,=\,4.0\,$\times$\,10$^6$\,\ls\\
$V_\infty$\,=\,400\,\kms\\

{And also} similar to \cite{2012MNRAS.423.1623G}:\\
$R_{\hbox{\rm core}}\,=60$\,\rs\\
$\dot{M}$\,=\,8.0\,$\times$\,10$^{-4}$\,\Msunyr\\
$L$\,=\,5.0\,$\times$\,10$^6$\,\ls\\
$V_\infty$\,=\,420\,\kms\\

The differences in the physical parameters are explained as follows.
The luminosity comes from infrared measurements - its precision (for a fixed distance) is primarily limited by a correction for flux not absorbed by dust, and a correction for the luminosity of the companion star. Due to the large mass-loss rate, the primary wind is optically thick, even in the visible. The core radius cannot be well determined - possible values range from 60 to several hundred \rs. Larger values are preferred if we assume that the \ion{He}{i} line emissions arise primarily in the wind-collision interface between the two stars as a consequence of the ionising flux of the secondary star. In  this work we adopt $R_{\hbox{\rm core}}\,=\,120$\,\rs.  

Models of the central star are riddled with systematic uncertainties arising from the following:

\begin{itemize}

\item Due to circumstellar extinction, we do not know the amount of reddening, nor the shape of the local extinction law. As a consequence, there is no absolute flux scale in the optical or UV.

\item  As discussed below, the circumstellar extinction alters the spectral appearance of the star (even with HST data). As a consequence, \citet{hillier2001}, did not model the ``intrinsic'' spectrum of the central star in the modelling of the HST spectrum obtained near periastron in December of 1998.

\item  The companion star influences the observed spectrum via its ionising radiation and the interaction of its wind with that of the primary  \citep{2012MNRAS.423.1623G, 2012ApJ...759L...2G, 2012ApJ...746L..18M}. At apastron, the companion star lies at a distance of $\sim$\,30\,au ($\sim$\,6500\,\rs) from the primary. The companion star carves a cavity in the primary's wind, and this cavity partially overlaps (in radius) with the line formation region in the primary wind. Its influence on the primary's spectrum is partially governed by the opening angle of the cavity, and hence the relative wind momenta of the two stars.  While the secondary is believed to be less luminous than the primary \citep[e.g.][]{MDF10_comp}, its ionising UV radiation field ($\lambda$\,$<$\,912\,{\AA}) is much stronger than that of the cooler primary.

\item Line emission may also arise from the wind collision zone.

\item The wind may be asymmetric \citep{SDG03_lat,vanboekel03,weigelt07}.

\end{itemize}

Despite these difficulties and uncertainties, the model of the primary star can still be used to glean insights into the nature of the coronagraph.

\subsection{CMFGEN studies of impact on the spectrum by different obstructing structures}
\label{cmfgenmodels}
 
\begin{figure}
 \centering
 {\includegraphics[viewport=100bp 10bp 570bp 470bp, clip, width=\linewidth]{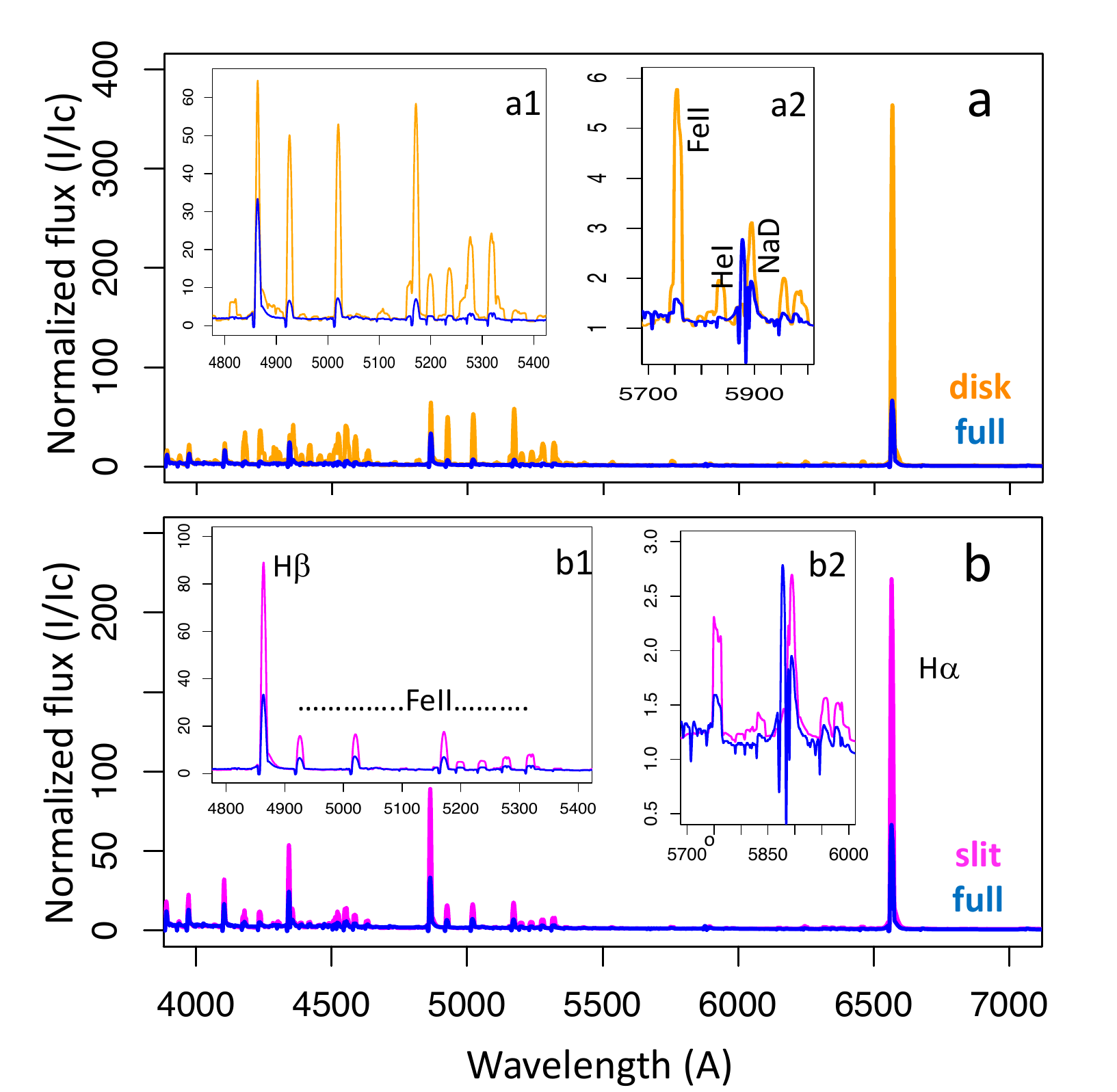}}
 \caption{{ Spectra from CMFGEN models with and without obstruction of the central star by a coronagraph}. 
 The {\it blue line} is the spectrum of the entire object (star plus wind)  {\it Panel a:} obstructing coronagraph as a circular disk centered on the star with radius $r$\,=\,11.5\,mas (26.45\,au). {\it Panel b:} extraction inside a slit of infinite length and 25\,mas wide, in which the star is placed at 2.5\,mas (5.75\,au) from the slit border.  Insets display zoomed views of spectral regions of interest.}
 \label{cmfgen_all}
\end{figure}

 {
Two models were constructed with the CMFGEN model \citep{hillier98} to assess the impact of an obstruction in front of the central star. First, we computed the unobscured optical spectrum of \ec\,A as shown  by a {\it blue line} in Fig.\,\ref{cmfgen_all}. The {\it orange line} in Fig.\,\ref{cmfgen_all}.a  shows  the spectrum integrated outside a circular disk of radius $r$\,$=$\,11.5\,mas, concentric to the star. In Fig.\,\ref{cmfgen_all}.b the {\it magenta line} shows the integrated flux from inside a slit of infinite length and 25\,mas wide. The slit is offset perpendicularly to its length so that the centre of the star is 2.5\,mas outside the slit. This second geometry is inspired by the emission structure observed by \citet{millour2020} and by \citet{falcke96} and by the spatial variability of the nebular ionisation analysed by Gull (in preparation). As expected, the coronagraph enhances the wind lines relative to the unimpeded view of the spectrum in both geometries, contributing to the majority of the continuum flux. Since most of the \ion{Fe}{ii}, and particularly the [\ion{Fe}{ii}]  emission lines are formed at larger radii than the Balmer lines, they show a greater increase in the EW absolute values. For the same reason, the intensity ratios of \ha to the upper members of the Balmer series also increase. 
}

In Fig.\,\ref{cmfgen_all}, excluding the Balmer lines, the strongest lines are from \ion{Fe}{ii}, while the weaker lines are due to [\ion{Fe}{ii}]. The \ion{He}{i} transition at $\lambda$\,5876{\AA} is much stronger in the full spectrum (blue curve) since the line forms close to the star because  Helium is mainly neutral in the outer wind.

There are numerous possibilities for the geometry and size of the coronagraph, including sharp and smooth boundaries. Our intent was not to fit the observed spectrum, which demands additional constraints from the observations, but to explore the general { EW trends} of an occulter in front of the star and its wind spectrum.

 In Table\,\ref{cmfgen1} we present the EW for a set of lines calculated in three CMFGEN model spectra shown in Fig.\,\ref{cmfgen_all}:  the unobscured view (full); a coronagraph occulting only the stellar core (disk); and a the occulter covering the whole field of view, except for a slit, which are compared to two observed spectra: one taken with FEROS/ESO in 1999.1, when the impact of the coronagraph was still large ($A_{\rm V}\,\sim\,4$\,mag), and one taken with NRES/LCOGT spectrograph (with instrumental setup similar to that of FEROS/ESO) in 2019.9, when the impact of the coronagraph had decreased to $A_{\rm V}\,\sim\,1$\,mag.
 The decrease in line strengths between these two dates follows the same trend in the spectrum changing from that of an occulted half stellar wind (slit) to the unocculted spectrum (full). This is clearly demonstrated in Table\,\ref{cmfgen1} which shows EW as a function of the models (columns 2, 3 and 4) and for the different epochs (columns 5 and 6).  

\setlength{\tabcolsep}{6.5pt}
\begin{table}
{
	\caption{Equivalent Width (in \AA) of representative lines extracted from CMFGEN models and observations.}
	\label{cmfgen1}
	\begin{tabular}{lrrrrr} 
    \hline				
Feature                    & full$^c$&disk$^b$  & slit$^a$ & 1999.1$^d$ & 2019.9$^d$ \\
\hline
\ha                        & $-$646 & $-$3675 & $-$2504  & $-$876 & $-$551 \\
H$\beta$                   & $-$131 &  $-$355 &  $-$444  & $-$139 & $-$121 \\
\hd                        &  $-$35 &  $-$81  &  $-$121  &  $-$28 &  $-$21 \\
\ion{Fe}{ii} 5317          &  $-$15 &  $-$178 & $-$67    &  $-$13 &   $-$6 \\
\hline
\end{tabular} \\
{{Notes:}
$^a$~extraction inside a slit aperture from $d$\,=\,2.5 to 27.5\,mas (5.75 to 63.25\,au) of the central star and infinite length;
$^b$~extraction outside a circle with $r$\,=\,11.5\,mas (26.45\,au);
$^c$~spectrum extracted over the entire field (star+wind);
$^d$~observations taken with FEROS/ESO (1999.9) and NRES/LCOGT (2019.9) instruments.}}
\end{table}
\setlength{\tabcolsep}{6pt}

 \subsection{Observational constraints for the position and shape of the coronagraph}
 \label{observational}

While the dusty occulter may be changing by tangential motion, irregular, clumped substructure and long-term dissipation, we describe a simple model to constrain its size. Assume the coronagraph  is circular, homogeneous, motionless, concentric with the primary star, and only its transparency is decreasing. The maximum size of the coronagraph can then be derived by using the projected distance of the Weigelt knots \citep{weigelt86} closest to the central star. While Weigelt\,B has faded, the two other clumps (C and D) have remained constant in flux, at the same phase of the orbital cycle, in the 1998-2004 period, in the optical and UV windows \citep{Gull16}. Unlike the central source, the Weigelt knots suffer very low extinction, $A_{\rm V}$\,$\sim$\,2.0\,$\pm$\,0.5\,mag  \citep{teodoro20}. Since the Weigelt C object was within $\approx$\,200\,mas (462\,au) from the central source in 1995  \citep{weigelt12}, we deduce that  the radius of the coronagraph must be smaller than this, assuming it to be spherical. The coronagraph cannot be located between  Weigelt C and the binary system since the high-excitation spectrum of Weigelt C has remained constant over the last two decades. { The coronagraph seems not to have the same nature as the Weigelt knot, because there are indication that it is more extended, covering the SE direction. In addition, the Weigelt knots are located at Homunculus equator, not to our LOS to the central binary.
}

An estimate of the minimum size along our LOS can be obtained from estimates of the circumstellar extinction. In the modelling by \citet{hillier2001}, the derived $V$-band circumstellar extinction was approximately 5 magnitudes in 1998, which agrees with \cite{damineli19}. With the physical model of the coronograph described above, and assuming the coronagraph has infinite optical depth, we obtain for a concentric circular configuration, a minimum radius of 60\,au (26\,mas). This is roughly the radius at which the electron scattering optical depth is 0.01. Since the spectrum extracted beyond this radius shows distinct differences from that of the primary star, the coronagraph must have some transparency, and thus it must have been larger than 60\,au in 1998.  However, its size is not well constrained to date. The flux increase observed since 1998 could be due to either an increase in transparency, a decrease in radius or a shift in the coronagraph location relative to our LOS.

Of course, the coronagraph does not have to be circular. \citet{falcke96} performed speckle polarimetry and resolved an obscuring $\sim$\,20\,$\times$\,60\,mas SW-NE dust bar in polarised light. Recently, \cite{millour2020} reported adaptive optics observations of \ec obtained with SPHERE at the VLT/ESO. The spatial resolution was 20\,mas ($\sim$\,47\,au), and observations were obtained in both  \ha and the nearby continuum. While difficult to interpret unambiguously, the images show structures that might be connected to the coronagraph. The stellar core is occulted by a dust lane running SW to NE and there is a bi-conical structure in the polarisation map running perpendicular to it (their Fig. 4). Further, the area NW of the central star is more transparent than the opposite side, consistent with markedly less extinction to Weigelt\,C and D.

\subsection{Consequences of the disappearance of the coronagraph}
\label{end}

The absolute EW of the eta1.079 feature is fading at a rate of $-$29\,{m\AA}\,yr$^{-1}$, so it will likely disappear in $\sim$\,2030. The uncertainty in this prediction is large because this feature is faint and the spectra are noisy, but it is consistent with the prediction for the coronagraph vanishing time-scale derived from the decreasing rate of the $A_{\rm V}$ extinction. 
 Following \citet{damineli19} (their Fig.\,18) this will happen in the time frame 2028-2036, depending on the foreground extinction, which is uncertain in the range $A_{\rm V}$\,=\,2.4-1.5\,mag.

The disappearance of the coronagraph is expected to have an impact also on the absorption lines of the spectrum components, since it will allow the light from the secondary star to ionise far-out regions of the circumstellar medium that were previously shielded from it { as we have shown for the narrow \ha\ absorption in the LH.  The details will be evaluated only after the size and position of the coronagraph are known.}

The correlation shown in Fig.\,\ref{AVDIB2} indicates that the eta1.079 carrier molecules are related to dust grains and both are being photo-dissociated by the intense stellar radiation field.  When the coronagraph has completely { vanished}, the stellar core will stop increasing in  brightness. The predicted magnitude of the stellar core at that time is $V$\,$\sim$\,2.5-3.5\,mag, which is a little brighter than it was in the 1600s{bf  (the amount of light reflected from the Homunculus is not very relevant here, because it was absent before the GE and now is fainter than the central star)}. The true luminosity variation depends on how much the bolometric correction (BC) changed since then, but unfortunately there were no historical reports on the colour of the star before the GE event. Let us assume that the change in BC was modest, which implies that the intrinsic luminosity did not change significantly or might have mildly increased. Since the GE ejected $>$\,15\,M$_\odot$ following \citet{morris99} and \citet{smith03} or $>$\,45\,M$_\odot$ following \citet{morris17} - using a full radiative transfer model with a far more chemically realistic dust library - the luminosity of the central star should have dropped by a large amount, unless it was compensated by some process. This favours the scenario of a merger in a triple system \citep{smith18, portegies15, hirai2020} over a massive wind ejection as the mechanism for the GE, since the primary star would have been mixed internally in a merger, bringing fresh fuel to its central core.

\section{Conclusions}
\label{sectionconclusions}

This work brings together various pieces of evidence indicating the presence of an obscuring structure in front of the binary system in \ec. Some of them were suspected as long as 30 years ago but now there are better observational data and modelling. 

The central stellar binary (\ec\,A\,+\,\ec\,B) has been brightening by many magnitudes for half a century, without major changes in the intrinsic spectrum of the primary star. This is not expected for a star which was already close to the Eddington limit at the beginning of this process. The long-term brightening has a circumstellar origin. The Homunculus reflection nebula has remained at an essentially constant brightness ($V$\,$\sim$\,5.48\,mag) while the central stellar core has brightened by several apparent magnitudes \citep{damineli19}. 

The main conclusions of this spectroscopic study are:

\begin{enumerate}[i]
    \item Spectral lines formed in the stellar wind have larger absolute EW in { direct than in reflected} light (Fig.\,\ref{directxreflected}).  

    \item The absolute EW values of the lines formed in the wind have decreased systematically from cycle to cycle  (Fig.\,\ref{FeII}.a). However, the Weigelt knots located at a  projected distance of a fraction of an arcsec  from the central star have maintained a constant brightness and ionisation level (Fig.\,\ref{FeII}.b).
    
    \item After correcting for the secular brightening, the  broad lines formed in the stellar wind exhibit a constant line profile (Fig.\,\ref{FeII}.c).
     
    \item Line profiles of the upper Balmer series members have been almost constant for the last 20 years, as compared to large variations in \ha, \ion{Fe}{ii} and [\ion{Fe}{ii}]. This is explained by the fact that these lines are formed at large radii in the stellar wind, as compared with the high-energy spectral lines, which are  mainly formed behind the coronagraph (Fig.\,\ref{hahdel}). The absolute EW value of \ha in direct light has decreased by $\approx$\,50\%  from 1985 to 2014 (Fig.\,\ref{Ha-Hd-long}). In contrast, the EW of  \ha observed in the FOS4 reflected light remained roughly constant over the time interval 2000-2010. The \ha EW in direct light converged to a value close to that from the FOS4 reflected light in 2014 and stalled afterwards (Fig.\,\ref{Ha-Hd-long}). EWs from both views remained at the same level during the last orbital cycle, suggesting that the coronagraph might have stopped vanishing. 
     Alternatively, the coronagraph may have fully { vanished}, and this has happened $\sim$\,10\,years earlier than previously predicted \citep{damineli19}. Spectra produced by CMFGEN modelling in the presence or not of an artificial coronagraph  of varying sizes, shapes and positions,  are in general agreement with observations (Fig.\,\ref{cmfgen_all} and Table\,\ref{cmfgen1})
    
    \item The constancy of the \hd and \ha EWs at the FOS4 reflected light reveals that the primary star in \ec is more stable than has often been claimed, for example by \citet{mehner12}. This indicates that whatever occurred during the GE, after $\sim$\,two centuries the star has become stable again - see (Figs.\,\ref{hahdel} and \ref{Ha-Hd-long}).
    
   \item The variability of the coronagraph is not completely smooth as seen in the short-lived event that occurred at the end of the 1990's. This could be due to irregularities in the transparency of a moving occulter, or a temporary variation in the rate of vanishing (Fig.\,\ref{figstructures}).

\item The variability curve of the \ha narrow absorption (``shell'') line (Fig.\,\ref{Hanar-historic}) has a complex pattern, indicating that it is produced by a number of structures that may be moving across our LOS. All of these structures produce an ionisation shadow in the LH at the position crossed by our LOS to the binary system. The non-periodic component of \ha narrow absorption (``shell'') in Fig.\,\ref{Hanar-historic} is produced by the coronagraph and it is weakening over  the long term, in line with other evidence.

{
\item In addition to the coronagraph, there is a second kind of  occulter that crosses our LOS to the binary system around periastron, when the ionising secondary star is on the back side. This occulter is the primary's  wind, which casts an ionisation shadow to the LH at our LOS close to periastron. It produces the periodic component displayed in the variability curve of the \ha narrow absorption.
}
 
\item The fact that the disappearance of the coronagraph impacts the strength of absorption lines originating in the LH indicates that it is located inside the LH. 

{
\item Our best guess is that the coronagraph is a sheet-like a circumbinary disk seen at some inclination to our LOS - the NW border of which runs over the central stellar object and is extended to the SE as suggested by the \citet{millour2020} and by \citet{falcke96}  high spatial resolution polarimetric maps. The permanent \ha\ shell absorption reported by \citet{boumis98} could be produced by the same disk in the Homunculus equator. Partial occultation to the SE direction is suggested by the slow EW fading of \ion{Fe}{ii}$\lambda$4585 reflected at FOS4. This evidence points to a 3D geometry of the occulter.
}

{
\item
We could not identify the origin of the absorption feature at $\lambda$10792\,\AA. It is unlikely to be formed in the primary's stellar wind, because there is no corresponding feature in any other spectral line of \ec. It does not seem to be produced by the same DIB known in the ISM, since it has positively displaced absorption components to our LOS ($+$\,45\,\kms) and ISM DIBs are produced at much colder temperatures. Because its decreasing EW is tightly correlated with the decreasing extinction, we believe it is located inside the coronagraph, produced by an unidentified process/molecule.
 }
\end{enumerate}

The different molecular absorptions, with a range of velocities and equilibrium temperatures observed by ALMA indicate that there is a contribution of absorbers from a range of distances from the central star in our direction.     
Regarding the end of the changes in the coronagraph, if it will continue up to the year 2030, the brightness of the central star will surpass that of the situation previous to the GE. If this happens, taking into account the large quantity of mass shed during the GE, the primary star should have been rejuvenated and increased its luminosity. { This would be in favour of the scenario of a binary merging in a triple stellar system.  However,  the question of how much such an internal mixing process could compensate for the lost mass can only be tackled by a specific model.}

Significant progress to map the coronagraph could be made with high spatial resolution ($\sim$\,mas) spectroscopy by sampling molecular absorption lines or resonant emission-lines.

\section*{Acknowledgements}
AD thanks to FAPESP for support award numbers 2011/51680-6 and 2019/02029-2. FN acknowledges FAPESP (2017/18191-8). AFJM is grateful for financial aid to NSERC (Canada). MFC was supported by NASA under award number 80GSFC21M0002. { We thank the referee A. Mehner for many constructive criticisms.}

\section*{Data availability}

The data underlying this article will be shared on reasonable request to the corresponding author.

\bibliographystyle{mnras}
%\bibliography{refs} % if your bibtex file is called example.bib

\label{lastpage}

\clearpage
%APPENDIX A ================================
\appendix
\section{Extended data tables}
\label{appendix}

\begin{table*}
	\centering									
	\caption{DIB1 and DIB2 Equivalent widths in normal and emission-line stars}
    \label{table-DIBs}
    \begin{tabular}{lccccl} 
\hline							
Object & EW $\lambda$10780 & $\sigma_{\rm{EW}}$ & EW $\lambda$10792	& $\sigma_{\rm{EW}}$ & Spectral Type\\
  & {(m\AA)}	& {(m\AA)}	& {(m\AA)}	& {(m\AA)}	&\\
\hline
$\beta$\,Monocerotis	&	72	&	10	&	66	&	10&B4Ve\\
$\delta$\,Scorpii	&	180	&	20	&	136	&	18&B0.3IVe	\\
$\beta$\,Lyrae	&	82	&	12	&	56	&	9&B8.5Ib-II	\\
HD\,169454	&	169	&	18	&	86	&	11&	A1V\\
AG\,Carinae	&	130	&	15	&	66	&	9&	WN11h-LBV\\
$\chi$\,Ophiuci	&	94	&	13	&	46	&	8& B2Vne	\\
HD\,168625	&	242	&	30	&	110	&	15&B6Iap-LBV	\\
HR\,Carinae	&	196	&	27	&	89	&	16&	LBV\\
HD\,168607	&	221	&	18	&	74	&	10&B9Iaep-LBV	\\
MWC\,314	&	267	&	30	&	87	&	12&B3Ibe-LBV	\\
P\,Cygni	&	91	&	16	&	29	&	7& LBV	\\
HD\,152236	&	61	&	16	&	19	&	7&B1Ia-0ek	\\
HD\,316285	&	387	&	100	&	90	&	20&B0Ieq-LBV	\\
WD1-W243	&	690	&	120	&	134	&	30&	LBV\\
\hline
\end{tabular} \\
\end{table*}

\begin{table*}
	\centering									
	\caption{Equivalent widths of the $\lambda$10792\,{\AA} absorption in eta Carinae observed at OPD/Coud\'e}
	\label{table-eta1.079}
	\begin{tabular}{ccccccccccc} 
    \hline							
Year	& EW &	$\sigma_{\rm{EW}}$ & Year	& EW &	$\sigma_{\rm{EW}}$ & Year	& EW &	$\sigma_{\rm{EW}}$ \\
 & {(m\AA)}&{(m\AA)}& & {(m\AA)}&{(m\AA)}& &{(m\AA)} &{(m\AA)} \\
\hline
1990.068	&	650	&	66	&	2001.446	&	550	&	52	&2010.177	&	490	&	47		\\
1991.398	&	650	&	100	&	2002.325	&	520	&	93	&2010.498	&	500	&	109		\\
1993.32	    &	630	&	68	&	2002.326	&	530	&	57	&2011.196	&	490	&	39		\\
1993.553	&	630	&	40	&	2002.331	&	520	&	45	&2011.53	&	440	&	45		\\
1995.36	    &	660	&	68	&	2002.547	&	540	&	35	&2011.531	&	500	&	52		\\
1996.075	&	660	&	45	&	2003.360    &	540	&	52	&2012.504	&	420	&	40		\\
1996.346	&	640	&	66	&	2004.502	&	520	&	103	&2013.537	&	450	&	44		\\
1997.381	&	630	&	75	&	2005.567	&	500	&	48	&2014.582	&	401	&	39	\\
1997.482	&	600	&	82	&	2006.199	&	530	&	55	&2014.585	&	398	&	57		\\
1998.054	&	610	&	58	&	2006.429	&	500	&	42	&2014.591	&	384	&	41		\\
1998.103	&	620	&	49	&	2006.637	&	530	&	99	&2014.593	&	400	&	36		\\
1998.284	&	620	&	90	&	2007.491	&	520	&	54	&2014.596	&	407	&	48		\\
1999.125	&	620	&	59	&	2007.494	&	530	&	85	&2016.486	&	387	&	75		\\
1999.316	&	540	&	84	&	2008.381	&	500	&	52	&2017.595	&	402	&	49		\\
2000.132	&	530	&	50	&	2009.131	&	510	&	88	&2018.417	&	380	&	42		\\
2000.945	&	530	&	44	&	2009.134	&	550	&	101	&2019.322	&	340	&	34	\\
2001.435	&	550	&	35	&	2009.301	&	480	&	38	&2019.542   &   360 & 32     \\
2001.438	&	540	&	57  &	2009.430    &	510	&	110	&2021.218    &  334 & 37	\\
\hline
\end{tabular} \\
\end{table*}

\begin{table}
	\centering									
	\caption{Equivalent widths in direct light for  \ha (first 10 rows). The full table is available online as supplementary material and in the CDS.}\label{table-direct-ha}
	\begin{tabular}{cccc} 
    \hline							
HJD	&	EW \ha	&	$\sigma_{\rm{EW}}$	&	Site	\\
 & {(\AA)}&{(\AA)}&  \\
\hline
49063.609	&	$-$1009	&	48	&	ESO/FEROS	\\
49065.609	&	$-$1013	&	51	&	ESO/FEROS	\\
49066.621	&	$-$996	&	49	&	ESO/FEROS	\\
49067.605	&	$-$1005	&	50	&	ESO/FEROS	\\
49068.609	&	$-$1001	&	52	&	ESO/FEROS	\\
49069.609	&	$-$1009	&	50	&	ESO/FEROS	\\
49071.613	&	$-$1006	&	50	&	ESO/FEROS	\\
49072.613	&	$-$962	&	52	&	ESO/FEROS	\\
49073.648	&	$-$1004	&	49	&	ESO/FEROS	\\
49074.625	&	$-$998	&	49	&	ESO/FEROS	\\
\hline
\end{tabular}
\end{table}

\begin{table}
	\centering									
	\caption{Equivalent widths in direct light for \ion{Fe}{ii}$\lambda$4585 (first 10 rows). The full table is available online as supplementary material and in the CDS.}\label{table-direct-fe2}
	\begin{tabular}{cccc} 
    \hline							
HJD	&	EW \ion{Fe}{ii}$\lambda$4585	&	$\sigma_{\rm{EW}}$	&	Site	\\
 & {(\AA)}&{(\AA)}&  \\
\hline
48844.4880	&	$-$16.1	&	1.8	&	ESO/FEROS	\\
49063.6090	&	$-$15.2	&	1.7	&	ESO/FEROS	\\
49067.6050	&	$-$16.0	&	1.7	&	ESO/FEROS	\\
49069.6090	&	$-$16.0	&	2.8	&	ESO/FEROS	\\
49071.6130	&	$-$16.1	&	1.5	&	ESO/FEROS	\\
49072.6130	&	$-$15.0	&	3.2	&	ESO/FEROS	\\
49073.6480	&	$-$15.3	&	1.3	&	ESO/FEROS	\\
49075.5740	&	$-$15.2	&	2.3	&	ESO/FEROS	\\
49076.6250	&	$-$15.0	&	2.7	&	ESO/FEROS	\\
49077.5900	&	$-$15.4	&	2.3	&	ESO/FEROS	\\
\hline
\end{tabular}
\end{table}

\begin{table}
	\centering									
	\caption{Equivalent widths in direct star for \hd (first 10 rows). The full table is available online as supplementary material and in the CDS.}\label{table-direct-hd}
	\begin{tabular}{cccc} 
    \hline							
HJD	&	EW \hd	&	$\sigma_{\rm{EW}}$	&	Site	\\
 & {(\AA)}&{(\AA)}&  \\
\hline
58502.451	&	$-$27.4	&	 1.0	&	LCO/NRES	\\
59039.234	&	$-$25.4	&	 2.0	&	LCO/NRES	\\
59051.232	&	$-$26.4	&	 0.6	&	LCO/NRES	\\
59152.602	&	$-$24.1	&	 1.4	&	LCO/NRES	\\
59155.600	&	$-$25.0	&	 1.2	&	LCO/NRES	\\
59174.562	&	$-$24.6	&	 0.8	&	LCO/NRES	\\
59206.513	&	$-$23.8	&	 1.1	&	LCO/NRES	\\
59210.463	&	$-$24.5	&	 0.9	&	LCO/NRES	\\
49063.609	&	$-$28.8	&	13.5	&	ESO/FEROS	\\
49065.609	&	$-$32.5	&	12.2	&	ESO/FEROS	\\

\hline
\end{tabular}
\end{table}

\setlength{\tabcolsep}{4pt}
\begin{table*}
	\centering									
	\caption{Equivalent widths of \ha\, \hd\ and  \ion{Fe}{ii}$\lambda$4585 from UVES observations extracted at FOS4 position.}\label{table-uves-fos4}
	\begin{tabular}{cccc|cccc|cccc} 
    \hline							
JD-2400000 & year	&	EW \hd	&	$\sigma_{\rm{EW}}$	&	JD-2400000 & year	&	EW \ion{Fe}{ii}$\lambda$4585	&	$\sigma_{\rm{EW}}$	&	JD-2400000 & year	&	EW \ha	&	$\sigma_{\rm{EW}}$	\\
 & & {(\AA)}&{(\AA)}& & & {(\AA)}&{(\AA)}& & & {(\AA)}&{(\AA)} \\
\hline
51533.832	&	1999.97	&	$-$22.00	&	0.61	&	51533.832	&	1999.97	&	$-$7.00	&	0.49	&	52616.833	&	2002.94	&	$-$521	&	8	\\
52634.854	&	2002.99	&	$-$20.00	&	0.62	&	52634.854	&	2002.99	&	$-$5.64	&	0.41	&	52639.852	&	2003.00	&	$-$560	&	9	\\
52639.840	&	2003.00	&	$-$21.00	&	0.57	&	52662.862	&	2003.06	&	$-$6.92	&	1.38	&	52642.832	&	2003.01	&	$-$555	&	9	\\
52662.862	&	2003.06	&	$-$19.60	&	0.61	&	52674.862	&	2003.09	&	$-$6.90	&	0.21	&	52658.814	&	2003.05	&	$-$530	&	9	\\
52662.862	&	2003.06	&	$-$19.53	&	0.61	&	52990.816	&	2003.96	&	$-$5.90	&	0.36	&	52662.848	&	2003.06	&	$-$532	&	9	\\
52968.807	&	2003.90	&	$-$20.04	&	0.73	&	53029.774	&	2004.07	&	$-$5.55	&	0.38	&	52674.848	&	2003.09	&	$-$530	&	9	\\
52990.816	&	2003.96	&	$-$19.93	&	0.70	&	53055.653	&	2004.14	&	$-$6.54	&	0.43	&	52684.657	&	2003.12	&	$-$518	&	9	\\
53029.774	&	2004.07	&	$-$20.33	&	0.67	&	53075.562	&	2004.19	&	$-$6.91	&	0.46	&	52990.803	&	2003.96	&	$-$564	&	6	\\
53055.653	&	2004.14	&	$-$21.32	&	0.67	&	53389.738	&	2005.05	&	$-$6.50	&	0.58	&	53029.763	&	2004.07	&	$-$540	&	7	\\
53075.562	&	2004.19	&	$-$20.56	&	0.93	&	53431.818	&	2005.17	&	$-$6.18	&	0.48	&	53075.703	&	2004.19	&	$-$564	&	9	\\
53389.738	&	2005.05	&	$-$19.08	&	0.82	&	53866.524	&	2006.36	&	$-$6.68	&	0.51	&	53349.850	&	2004.94	&	$-$535	&	11	\\
53431.818	&	2005.17	&	$-$19.50	&	0.66	&	53912.550	&	2006.48	&	$-$6.84	&	0.55	&	53389.709	&	2005.05	&	$-$540	&	14	\\
53866.524	&	2006.36	&	$-$19.88	&	0.72	&	54535.755	&	2008.19	&	$-$6.67	&	0.41	&	53389.725	&	2005.05	&	$-$541	&	8	\\
53912.550	&	2006.48	&	$-$21.18	&	0.76	&	54554.759	&	2008.24	&	$-$5.62	&	0.44	&	53431.805	&	2005.17	&	$-$527	&	26	\\
54513.829	&	2008.13	&	$-$21.32	&	1.24	&	54567.535	&	2008.28	&	$-$5.20	&	0.66	&	53912.538	&	2006.48	&	$-$540	&	13	\\
54535.755	&	2008.19	&	$-$20.10	&	0.58	&	54583.543	&	2008.32	&	$-$5.70	&	0.41	&	54475.731	&	2008.03	&	$-$540	&	12	\\
54554.759	&	2008.24	&	$-$21.00	&	0.67	&	54599.492	&	2008.36	&	$-$5.24	&	0.44	&	54554.740	&	2008.24	&	$-$523	&	13	\\
54567.535	&	2008.28	&	$-$21.79	&	0.70	&	54629.491	&	2008.45	&	$-$5.27	&	0.38	&	54567.517	&	2008.28	&	$-$559	&	13	\\
54583.543	&	2008.32	&	$-$20.88	&	0.64	&	54657.552	&	2008.52	&	$-$5.35	&	0.47	&	54583.529	&	2008.32	&	$-$543	&	14	\\
54599.492	&	2008.36	&	$-$20.46	&	0.63	&	54999.565	&	2009.46	&	$-$4.34	&	0.39	&	54599.475	&	2008.36	&	$-$552	&	14	\\
54629.491	&	2008.45	&	$-$20.08	&	0.67	&	55013.483	&	2009.50	&	$-$4.86	&	0.34	&	54616.536	&	2008.41	&	$-$547	&	13	\\
54657.552	&	2008.52	&	$-$20.20	&	0.66	&	56620.846	&	2013.90	&	$-$4.00	&	0.71	&	54629.504	&	2008.45	&	$-$529	&	13	\\
54999.565	&	2009.46	&	$-$20.05	&	0.67	&	56650.835	&	2013.98	&	$-$4.31	&	0.71	&	56999.838	&	2014.94	&	$-$485	&	36	\\
55013.483	&	2009.50	&	$-$20.00	&	0.44	&	56684.722	&	2014.07	&	$-$4.05	&	0.61	&	56999.841	&	2014.94	&	$-$487	&	12	\\
56620.846	&	2013.90	&	$-$21.00	&	1.26	&	56712.686	&	2014.15	&	$-$4.28	&	0.74	&	57012.805	&	2014.97	&	$-$505	&	13	\\
57043.865	&	2015.06	&	$-$19.08	&	1.39	&	57043.865	&	2015.06	&	$-$3.68	&	0.82	&	57027.837	&	2015.01	&	$-$497	&	27	\\
57047.796	&	2015.07	&	$-$19.63	&	0.99	&	57047.796	&	2015.07	&	$-$3.87	&	0.59	&	57047.781	&	2015.07	&	$-$531	&	13	\\
\hline
\end{tabular}
\end{table*}
\setlength{\tabcolsep}{6pt}

\end{document}